\documentclass[a4paper,11pt]{article}
\pdfoutput=1 %

\usepackage{jheppub} 
\usepackage{amsmath,amssymb,graphicx,float,slashed,xcolor,multicol}
\usepackage{tabularx}
\usepackage{url}
\usepackage{footmisc}
\usepackage{amsfonts}
\usepackage{cancel}
\usepackage{color}
\usepackage{multirow} 
\usepackage{pifont}
\usepackage{epstopdf}
\usepackage{comment}
\usepackage{booktabs} 
\usepackage{natbib}
\usepackage{array}
\usepackage{mathrsfs}
\usepackage[toc,page]{appendix}
\usepackage{mathtools}
\usepackage{romannum}
\usepackage{ulem}
\usepackage{bbold}
\usepackage{enumitem}
\usepackage{multirow}
\usepackage{cleveref}
\usepackage{caption,subcaption}
\usepackage{float}
\usepackage{multirow}
\usepackage{upgreek}
\usepackage[toc,page]{appendix}
\usepackage{romannum}
\allowdisplaybreaks
\usepackage{bbm}                
\usepackage{xspace}				

\usepackage{tikz}
\usetikzlibrary{arrows,shapes}
\usetikzlibrary{trees}
\usetikzlibrary{matrix} 
\usetikzlibrary{positioning}				
\usetikzlibrary{calc,through}				
\usetikzlibrary{decorations.pathreplacing}  
\usepackage{pgffor}							
\usetikzlibrary{decorations.pathmorphing}	
\usetikzlibrary{decorations.markings}
\tikzset{
	vector/.style={decorate, decoration={snake}, draw},
	provector/.style={decorate, decoration={snake,amplitude=2.5pt}, draw},
	antivector/.style={decorate, decoration={snake,amplitude=-2.5pt}, draw},
	fermion/.style={draw=black, postaction={decorate},
		decoration={markings,mark=at position .55 with {\arrow[draw=black]{>}}}},
	fermionbar/.style={draw=black, postaction={decorate},
		decoration={markings,mark=at position .55 with {\arrow[draw=black]{<}}}},
	fermionnoarrow/.style={draw=black},
	gluon/.style={decorate, draw=black,
		decoration={coil,amplitude=4pt, segment length=5pt}},
	scalar/.style={dashed,draw=black, postaction={decorate},
		decoration={markings,mark=at position .55 with {\arrow[draw=black]{>}}}},
	scalarbar/.style={dashed,draw=black, postaction={decorate},
		decoration={markings,mark=at position .55 with {\arrow[draw=black]{<}}}},
	scalarnoarrow/.style={dashed,draw=black},
	electron/.style={draw=black, postaction={decorate},
		decoration={markings,mark=at position .55 with {\arrow[draw=black]{>}}}},
	bigvector/.style={decorate, decoration={snake,amplitude=4pt}, draw},
}

\tikzset{>=latex} 

\usepackage{tikz-feynman}
\makeatletter
\tikzfeynmanset{compat=\tikzfeynman@version@major.\tikzfeynman@version@minor.\tikzfeynman@version@patch}
\makeatother

\tikzstyle{block} = [draw, rectangle, 
minimum height=3em, minimum width=6em]

\newcommand*{\rom}[1]{\expandafter\@slowromancap\romannumeral #1@}

\def\lag{\mathscr{L}}

\def\beq{\begin{equation}}
\def\eeq{\end{equation}}
\def\beqa{\begin{eqnarray}}
\def\eeqa{\end{eqnarray}}
\title{Scalar and tensor neutrino interactions}
\author[a]{Tao Han,}
\author[b]{Jiajun Liao,}
\author[a]{Hongkai Liu,}
\author[c]{Danny Marfatia}
\affiliation[a]{Department of Physics and Astronomy, University of Pittsburgh, Pittsburgh, PA 15260, USA }
\affiliation[b]{School of Physics, Sun Yat-sen University, Guangzhou, 510275, China}
\affiliation[c]{Department of Physics and Astronomy, University of Hawaii at Manoa, Honolulu, HI 96822, USA}
\emailAdd{than@pitt.edu}
\emailAdd{liaojiajun@mail.sysu.edu.cn}
\emailAdd{hol42@pitt.edu}
\emailAdd{dmarf8@hawaii.edu}

\preprint{
\begin{flushright}
PITT-PACC-2003
\end{flushright}
}

\abstract{ We constrain general Dirac neutrino interactions based on the Standard Model Effective Field Theory framework extended with right-handed neutrinos $N$ (SMNEFT) using deep inelastic and coherent elastic neutrino scattering, nuclear beta decay, and meson decay data, and high energy electron-proton and proton-proton collider data. We compute the one-loop anomalous dimensions of the low-energy effective field theory (LEFT) below the electroweak scale and of SMNEFT above the electroweak scale. The tree-level matching between LEFT and SMNEFT is performed at the electroweak scale. Currently, the most stringent limits on scalar and tensor interactions arise from pseudoscalar meson decays and the LHC measurements at the per mille level. In the future, the upcoming High-Luminosity LHC (HL-LHC) has the potential to reach  the $10^{-4}$ level and LHeC can play an important role under certain theoretical assumptions.}

\begin{document}

\titlepage

\maketitle



\flushbottom

\section{Introduction}
\label{sec:intro}
Flavor oscillations between massive neutrinos is a firmly established phenomenon that cannot be accounted for by the Standard Model (SM)~\cite{Weinberg:1967tq,Glashow:1961tr, Salam:1968rm}, thus strongly motivating physics beyond the SM (BSM) associated with the neutrino sector~\cite{Barger:2003qi, GonzalezGarcia:2007ib}. 
The absence of BSM signals at the LHC and other low energy experiments suggests that either the new physics scale is beyond the kinematic reach of our high energy experiments or that it is of a more subtle incarnation. It is therefore prudent to guide our search for new physics as generally as possible without theoretical prejudice.

Generically, departures from the SM at energies below a new physics scale 
can be described by a model-independent Effective Field Theory (EFT) after integrating out the heavy degrees of freedom in the new physics sector.
%
Such an effective Lagrangian was first constructed by Fermi for the 4-fermion contact interaction involving a neutrino in nuclear $\beta$-decay. In the SM language, the contact interaction is a result of integrating out a heavy particle, the electroweak $W$-boson. To account for 
nonstandard interactions (NSI) of neutrinos, Wolfenstein 
proposed 4-fermion interactions with general couplings~\cite{Wolfenstein:1977ue}, that have helped understand matter effects in solar neutrino oscillation experiments. Much theoretical and experimental effort has been made to search for potential new physics along the lines of neutrino NSI; for reviews see Refs.~\cite{Ohlsson:2012kf,Miranda:2015dra,Farzan:2017xzy}.
To explore new physics near or above the electroweak scale, it is appropriate to adopt an EFT, respecting the full SM gauge symmetry 
with the SM field content, the so-called Standard Model Effective Field Theory (SMEFT)~\cite{Grzadkowski:2010es,Henning:2014wua,Brivio:2017vri}. 
Neutrino NSIs at low energies can be obtained in this framework by integrating out the heavy particles $t,W^\pm,Z$ and the Higgs boson $h$. 
%
%
%

The addition of right-handed neutrino states to the SM is the simplest extension to account for neutrino oscillations. In this article, we adopt this framework including three right-handed neutrino states $N$ that are lighter than a keV. Naturally, they are ``sterile neutrinos'' with no SM gauge charges. 
However, we do not specify their possible Majorana mass terms. We restrict our study to the case in which the left-handed neutrino states are Dirac in nature. Then, 
new flavor-conserving neutral current vector and tensor interactions are possible.
The general Standard Model Effective Field Theory extended with right-handed neutrinos (SMNEFT) has been presented in Ref.~\cite{Liao:2016qyd, Bischer:2019ttk}. We follow this well motivated formalism.
A framework for model-independent General Neutrino Interactions (GNI) below the electroweak scale has been also constructed in Ref.~\cite{Lindner:2016wff}. All operators of scalar, pseudoscalar, vector, axial vector and tensor interactions of neutrinos with SM fermions are included, leading to potentially rich phenomenology. We note that scalar and tensor GNI operators cannot be embedded in SMEFT at the dimension-six level, but are present in SMNEFT. 

There are well-motivated ultraviolet complete models that introduce SM singlet right-handed neutrinos. $U(1)_{B-L}$ extensions of the SM generate vector interactions between right-handed neutrinos and SM particles. In models with left-right symmetry~\cite{Pati:1974yy,Mohapatra:1974gc,Senjanovic:1975rk}, such interactions are generated after left-right symmetry breaking. Leptoquark models~\cite{Pati:1973uk,Bischer:2019ttk}  generate various neutrino-quark interactions. Each of these leads to model-dependent phenomenology if the new states are kinematically accessible, which we will not explore further. Instead, we focus on generic features of Dirac neutrino interactions after integrating out the heavy degrees of freedom.

In this article, we set out to examine new scalar and tensor neutrino-quark interactions using SMNEFT. We present constraints on the operators from current low-energy data including neutrino mass bounds, meson and nuclear decays, neutrino deep inelastic scattering, and coherent elastic neutrino-nucleus scattering (CE$\nu$NS), and from current high-energy data from the LHC. We also explore the potential sensitivities for future  CE$\nu$NS, LHeC, and LHC experiments. 
Since the physical processes we consider span a vast range of energies, the renormalization group running and matching effects for the relevant Wilson coefficients at different energy scales  have been properly taken into account in our analyses; this procedure was carried out for vector NSI in Ref.~\cite{Terol-Calvo:2019vck}.

The rest of the paper is organized as follows. In section~\ref{sec:formalism} we provide an overview of the theoretical formalism and emphasize the importance of matching the Wilson coefficients by renormalization group running at different scales. In section~\ref{sec:low}, we present the current constraints and future sensitivities from a wealth of low energy experiments. In section~\ref{sec:high}, we study the current and projected bounds from high-energy proton-proton colliders and electron-proton colliders. We summarize our results in section~\ref{sec:sum}.
Some details of the renormalization group running are given in appendix~\ref{RGrunning}.

\section{Theoretical formalism}
\label{sec:formalism}
The 4-fermion neutrino interactions via the SM neutral and charged currents at the leading order (LO), after integrating out the $Z$ and $W^\pm$ propagators, are
\beqa 
\lag^{NC}_{\text{SM}} &=&-\frac{G_F}{\sqrt{2}} \delta^{\alpha\beta}\delta^{\gamma\delta} [\overline{\nu}_\alpha \gamma_\mu (1-\gamma_5) \nu_\beta ] [g_{L,f}\overline{f}_\gamma \gamma^\mu(1-\gamma_5)f_\delta + g_{R,f}\overline{f}_\gamma \gamma^\mu(1+\gamma_5)f_\delta]\,, 
\label{eq:NC}\\
\lag^{CC}_{\text{SM}}  &=& -\frac{G_F}{\sqrt{2}}V^*_{\delta\gamma} \delta^{\alpha\beta} [\overline{\nu}_\alpha \gamma_\mu (1-\gamma_5) \ell_\beta] [\overline{d}_\gamma\gamma^\mu(1-\gamma_5)u_\delta] + \text{h.c.}\,, 
\label{eq:CC}
\eeqa
where the Fermi constant  $G_F/\sqrt 2 = (2v^2)^{-1}=g^2/8M_W^2$. $f$ denotes quarks and charged leptons, $V$ is the CKM quark-mixing matrix,  $g$ is the SM SU(2)$_{\rm L}$ gauge coupling, $M_W$ is the mass of the $W$ boson given by the Higgs vacuum expectation value $gv/2$, and $\alpha,\;\beta,\;\gamma,\;\text{and}\;\delta$ are flavor indices. The chiral couplings $g_{L,f}$ and $g_{R,f}$ are defined as
\beq
g_{L,f} = T^3_f - Q_f\sin^2\theta_W\,,\quad g_{R,f} =  - Q_f\sin^2\theta_W\,,
\label{eq:eff_coup}
\eeq
where $Q_f$ is the fermion's charge in units of $+e$. We choose a flavor basis such that the down-type quark and the lepton Yukawa matrices are diagonal for convenience of calculation. The transformation for the up-type quarks between the flavor (primed) and mass basis (unprimed) reads
\beq
u^{\prime}_{L,\alpha} = V_{\alpha\beta}^\dagger\; u_{L,\beta}\,.
\eeq
In the spirit of EFT, the theory is valid only at low energies, $E \ll M_W$. 
	Note that NSI~\cite{Ohlsson:2012kf,Miranda:2015dra,Farzan:2017xzy}  are of the same form as Eqs.~(\ref{eq:NC}) and (\ref{eq:CC}), but the scale and couplings are free parameters to reflect the unknown nature of new physics.

\subsection{LEFT}

Going beyond the SM, the full list of dimension-six four-fermion SMNEFT operators, which include the left-handed and right-handed neutrino states, are given in Ref.~\cite{Bischer:2019ttk}. 
Integrating out $W^\pm, Z, h$ and $t$ leads to the low-energy effective field theory (LEFT), respecting $SU(3)_C\times U(1)_Q$. 
%
%
GNI via neutral and charged currents, containing scalar, pseudoscalar, vector, axial vector and tensor terms at dimension-six level can be parameterized as
\beqa
\label{eq:lowNC}
\lag^{\text{NC}}_{\text{LEFT}} &\supset& -\frac{G_F}{\sqrt{2}}\sum_{j=1}^{10} (\overset{(\sim)}{\epsilon}_{j,f})^{\alpha\beta\gamma\delta}(\overline{\nu}_\alpha O_j \nu_\beta)(\overline{f}_\gamma O^\prime_j f_\delta)\,, \\
\lag^{\text{CC}}_{\text{LEFT}} &\supset& -\frac{G_F V^*_{\delta\gamma}}{\sqrt{2}}\sum_{j=1}^{10} (\overset{(\sim)}{\epsilon}_{j,du})^{\alpha\beta\gamma\delta}(\overline{\nu}_\alpha O_j \ell_\beta)(\overline{d}_\gamma O^\prime_j u_\delta) + \text{h.c.}\,,
\label{eq:lowCC}
\eeqa
where the operators $O_j, O_j^\prime$ and parameters $\overset{(\sim)}{\epsilon}$ are listed in Table~\ref{Table: eps}. The Dirac spinor $\nu_\alpha=(\nu_{L\alpha}, N_\alpha)^T$, and $u$ and $d$ indicate the mass eigenstates of up- and down-type quarks, respectively.\footnote{Our analysis can also be applied to Majorana neutrinos
with the neutrino bilinears in Eqs.~(\ref{eq:NC}) and~(\ref{eq:lowNC}) replaced by $\overline{N}_{M\alpha} O \nu_{M\beta}$,  where the Majorana spinors are $\nu_M= (\nu_L, \nu_L^c)^T$, $N_M=(N^c, N)^T$.}
Hermiticity of the Lagrangian requires the scalar and tensor effective couplings to satisfy 
\beq
\begin{split}
&\epsilon^{\alpha\beta\gamma\delta}_{S,f} = (\tilde{\epsilon}_{S,f}^{\beta\alpha\delta\gamma})^*,\quad \epsilon^{\alpha\beta\gamma\delta}_{P,f} = -(\tilde{\epsilon}_{P,f}^{\beta\alpha\delta\gamma})^*,\quad \epsilon^{\alpha\beta\gamma\delta}_{T,f} = (\tilde{\epsilon}_{T,f}^{\beta\alpha\delta\gamma})^*\,,\\
&\epsilon^{\alpha\beta\gamma\delta}_{S,du} = (\tilde{\epsilon}_{S,ud}^{\beta\alpha\delta\gamma})^*,\quad \epsilon^{\alpha\beta\gamma\delta}_{P,du} = -(\tilde{\epsilon}_{P,ud}^{\beta\alpha\delta\gamma})^*,\quad \epsilon^{\alpha\beta\gamma\delta}_{T,du} = (\tilde{\epsilon}_{T,ud}^{\beta\alpha\delta\gamma})^*\,.\\
\end{split}
\label{Eq:hermitian}
\eeq
\begingroup
\setlength{\tabcolsep}{20pt} 
\renewcommand{\arraystretch}{1.} 
\begin{table}
	\centering
	\begin{tabular}{c  c  c  c}
		\toprule
		$j$  &  $\overset{(\sim)}{\epsilon}_{j}$  &  $O_j$   &  $O^\prime_j$ \\
		\midrule
		1  & $\epsilon_L$ & $\gamma_\mu (\mathbb{1}-\gamma^5)$ & $\gamma_\mu (\mathbb{1}-\gamma^5)$  \\
		
		2 & $\tilde{\epsilon}_L$ & $\gamma_\mu (\mathbb{1}+\gamma^5)$ &$\gamma_\mu (\mathbb{1}-\gamma^5)$  \\
		
		3 & $\epsilon_R$ & $\gamma_\mu (\mathbb{1}-\gamma^5)$ & $\gamma_\mu (\mathbb{1}+\gamma^5)$  \\
		
		4 & $\tilde{\epsilon}_R$ & $\gamma_\mu (\mathbb{1}+\gamma^5)$ & $\gamma_\mu (\mathbb{1}+\gamma^5)$  \\
		
		5 & $\epsilon_S$ & $\mathbb{1} - \gamma^5$ & $\mathbb{1}$  \\
		
		6 & $\tilde{\epsilon}_S$ & $\mathbb{1} + \gamma^5$ & $\mathbb{1}$  \\
		
		7 & $-\epsilon_P$ & $\mathbb{1} - \gamma^5$ & $\gamma^5$  \\
		
		8 & $-\tilde{\epsilon}_P$ & $\mathbb{1} + \gamma^5$ & $\gamma^5$  \\
		
		9 & $\epsilon_T$ & $\sigma_{\mu\nu} (\mathbb{1} - \gamma^5)$ & $\sigma_{\mu\nu} (\mathbb{1} - \gamma^5)$  \\
		
		10 & $\tilde{\epsilon}_T$ & $\sigma_{\mu\nu} (\mathbb{1} + \gamma^5)$ & $\sigma_{\mu\nu} (\mathbb{1} + \gamma^5)$  \\
		\toprule
	\end{tabular}
	\caption{Effective coupling constants and operators.}
	\label{Table: eps}
\end{table}
\endgroup
If the BSM new physics scale is $\Lambda$ with a typical tree-level coupling $\kappa$, then parametrically $\overset{(\sim)}{\epsilon} \sim \kappa^2 v^2/\Lambda^2$. Note that the operators with $j=1,3$ are the familiar NSI terms, and are a subset of SMEFT.

\subsection{SMNEFT}
\label{sec:SMNEFT}
Of the dim-6 4-fermion SMNEFT operators related to scalar and tensor GNI, the three chirality-flipping operators that couple to quarks are 
\begin{enumerate}
	\item $O^{\alpha\beta\gamma\delta}_{NLQu} =  (\overline{N}_{\alpha} L_{\beta}^j)(\overline{Q}^j_\gamma u_\delta)$\,,
	\item $O^{\alpha\beta\gamma\delta}_{NLdQ} = (\overline{N}_{\alpha} L^j_{\beta})\epsilon_{jk}(\overline{d}_\gamma Q^k_\delta)$\,,
	\item $O^{\prime\alpha\beta\gamma\delta}_{NLdQ} = (\overline{N}_{\alpha}\sigma_{\mu\nu} L^j_{\beta})\epsilon_{jk}(\overline{d}_\gamma \sigma^{\mu\nu} Q^k_\delta)$\,,
\end{enumerate}
where the fields are written in two-component spinors. $L$ and $Q$ are the left-handed lepton and quark doublet, respectively, and $N$ is the right-handed neutrino state. Here, $\sigma^{\mu\nu} = \frac{i}{2}[\sigma^\mu\overline{\sigma}^\nu - \sigma^\nu\overline{\sigma}^\mu]$, with $\sigma^\mu=(\mathbb{1}, \vec{\mathbf{\sigma}})$ and $\bar{\sigma}^\mu=(\mathbb{1}, -\vec{\mathbf{\sigma}})$. We do not consider other dim-6 4-fermion SMNEFT operators since they lead to nonstandard charged lepton interactions and are therefore strongly constrained~\cite{Bischer:2019ttk}.
We can write the effective Lagrangian as
\beq
\lag_{\text{eff}} = \lag_{\text{SM}}+2\sqrt{2} G_F [C_{NLdQ} O_{NLdQ} +  C_{NLQu} O_{NLQu} +  C^\prime_{NLdQ} O^\prime_{NLdQ}]\,,
\eeq
where the flavor indices are omitted for simplicity. In the same spirit of power counting as in the last subsection, the Wilson coefficients (WCs) have the general dependence $C \sim \kappa^2 v^2 / \Lambda^2$.
For instance, $C \sim {\cal O}(10^{-4})$ if $\Lambda \sim 10$~TeV and $\kappa \sim 1$.

To jointly interpret the results of experiments at very different energy scales, a consistent theoretical framework is needed. LEFT and SMNEFT are the language we use to describe the physics below and above the electroweak scale $v$, respectively. The renormalization group (RG) running below and above the electroweak scale makes it possible to directly compare low-energy and high-energy probes. 
Leading-order (LO) matching between these two EFTs is performed at the electroweak scale. 

\subsection{RG running and matching}
 Since we will use both low-energy neutrino scattering experiments and high-energy colliders to constrain these Wilson coefficients (WCs), renormalization group (RG) running and matching have to be implemented. We perform leading-order (LO) matching of these two EFTs at the eletroweak scale:
  \beq
\begin{split}
	&\epsilon^{\alpha\beta\gamma\delta}_{S,d} = -C_{NLdQ}^{\alpha\beta\gamma\delta}\,, \quad \epsilon^{\alpha\beta\gamma\delta}_{S,u} = -V_{\gamma\rho}C_{NLQu}^{\alpha\beta\rho\delta} \,,\\
	&\epsilon^{\alpha\beta\gamma\delta}_{P,d} = -C_{NLdQ}^{\alpha\beta\gamma\delta}\,, \quad \epsilon^{\alpha\beta\gamma\delta}_{P,u} = V_{\gamma\rho}C_{NLQu}^{\alpha\beta\rho\delta}\,,\\
	&\epsilon^{\alpha\beta\gamma\delta}_{T,d} = -C_{NLdQ}^{\prime\alpha\beta\gamma\delta}\,, \quad \epsilon^{\alpha\beta\gamma\delta}_{T,u} =0\,,\\
	& \epsilon^{\alpha\beta\gamma\delta}_{S,du}  = \frac{1}{V_{\delta\gamma}^*}(C_{NLdQ}^{\alpha\beta\gamma\rho}V_{\rho\delta}^\dagger - C_{NLQu}^{\alpha\beta\gamma\delta})\,, \quad \epsilon^{\alpha\beta\gamma\delta}_{P,du} = \frac{1}{V_{\delta\gamma}^*}(C_{NLdQ}^{\alpha\beta\gamma\rho}V_{\rho\delta}^\dagger+C_{NLQu}^{\alpha\beta\gamma\delta})\,,\\
	&\epsilon^{\alpha\beta\gamma\delta}_{T,du} =\frac{1}{V_{\delta\gamma}^*}C_{NLdQ}^{\prime\alpha\beta\gamma\rho}V_{\rho\delta}^\dagger\,.
	\label{match}
\end{split}
\eeq
 As we run down, both neutral and charged current WCs are induced by each of the three SMNEFT operators. Therefore the currents are not independent of each other. Their relations at the electroweak scale are
 \beq
\begin{split}
	&\epsilon^{\alpha\beta\gamma\delta}_{S,du} =\frac{1}{V^*_{\delta\gamma}}( V^\dagger_{\gamma\rho}\epsilon^{\alpha\beta\rho\delta}_{S,u}-\epsilon^{\alpha\beta\gamma\rho}_{S,d}V^\dagger_{\rho\delta})\,,\quad
	\epsilon^{\alpha\beta\gamma\delta}_{P,d} = \epsilon^{\alpha\beta\gamma\delta}_{S,d}\,,\\
	&\epsilon^{\alpha\beta\gamma\delta}_{P,du} = -\frac{1}{V^*_{\delta\gamma}}(\epsilon^{\alpha\beta\gamma\rho}_{S,d}V^\dagger_{\rho\delta} +V^\dagger_{\gamma\rho}\epsilon^{\alpha\beta\rho\delta}_{S,u})\,, \quad
	\epsilon^{\alpha\beta\gamma\delta}_{P,u} =-\epsilon^{\alpha\beta\gamma\delta}_{S,u}\,,\\ 
	&\epsilon^{\alpha\beta\gamma\delta}_{T,du} = -\frac{1}{V_{\delta\gamma}^*}\epsilon^{\alpha\beta\gamma\rho}_{T,d}V^\dagger_{\rho\delta}\,.
\end{split}
\label{eq:NCCC}
\eeq 
Note that $1/{V^*_{\delta\gamma}}$ in Eqs.~(\ref{match}) and (\ref{eq:NCCC}) is a number without summation indices.
 We have performed the RG running above and below the weak scale, the details of which are described in Appendix~\ref{RGrunning}. The RG equations are run from 2~GeV to 1~TeV, which is the typical LHC scale. Eventually we place bounds on the SMNEFT WCs at 1~TeV. The anomalous dimension matrices we calculated at the one-loop level are
\beq
\begin{split}
	\mu\frac{d}{d\mu}\begin{pmatrix}
		C_{NLQu}\\
		C_{NLdQ}  \\
		C^\prime_{NLdQ} \\
	\end{pmatrix}_{(\mu)} &= [ \frac{\alpha_1(\mu)}{2\pi}\begin{pmatrix}
		-1/3 & 0 & 0 \\
		0 & 1/6 & -1\\
		0 &  -1/48 & -5/9\\
	\end{pmatrix} + \frac{\alpha_2(\mu)}{2\pi}\begin{pmatrix}
		0 & 0 & 0 \\
		0 & 0 & 9\\
		0 &  3/16 & -3/2\\
	\end{pmatrix}\\
	&+\frac{\alpha_3(\mu)}{2\pi}\begin{pmatrix}
		-4 & 0 & 0 \\
		0 & -4 &  0\\
		0 &  0 & 4/3\\
	\end{pmatrix} ] \begin{pmatrix}
		C_{NLQu}\\
		C_{NLdQ}\\
		C^\prime_{NLdQ}\\
	\end{pmatrix}_{(\mu)},
\end{split}
\label{eq:adm_c}
\eeq
\beq
\begin{split}
	\mu\frac{d}{d\mu}\begin{pmatrix}
	\epsilon_{S,du}\\
	\epsilon_{P,du}\\
	\epsilon_{T,du}\\
	\end{pmatrix}_{(\mu)} &= [\frac{\alpha_e(\mu)}{2\pi}\begin{pmatrix}
		2/3 & 0 & 4 \\
		0 & 2/3 & 4\\
		1/24 &  1/24 & -20/9\\
	\end{pmatrix}
	+\frac{\alpha_3(\mu)}{2\pi}\begin{pmatrix}
		-4 & 0 & 0 \\
		0 & -4 &  0\\
		0 &  0 & 4/3\\
	\end{pmatrix}] \begin{pmatrix}
		\epsilon_{S,du}\\
		\epsilon_{P,du}\\
		\epsilon_{T,du}\\
	\end{pmatrix}_{(\mu)},
\end{split}
\label{eq:adm_eud}
\eeq
\beq
\begin{split}
	\mu\frac{d}{d\mu}\begin{pmatrix}
		\epsilon_{S,d}\\
		\epsilon_{P,d}\\
		\epsilon_{T,d}\\
	\end{pmatrix}_{(\mu)} &= [\frac{\alpha_e(\mu)}{2\pi}\begin{pmatrix}
		-1/9 & 0 & 0 \\
		0 & -1/9 & 0\\
		0 &  0 & 5/36\\
	\end{pmatrix}
	+\frac{\alpha_3(\mu)}{2\pi}\begin{pmatrix}
		-4 & 0 & 0 \\
		0 & -4 &  0\\
		0 &  0 & 4/3\\
	\end{pmatrix}] \begin{pmatrix}
		\epsilon_{S,d}\\
		\epsilon_{P,d}\\
		\epsilon_{T,d}\\
	\end{pmatrix}_{(\mu)},
\end{split}
\label{eq:adm_ed}
\eeq
\beq
\begin{split}
	\mu\frac{d}{d\mu}\begin{pmatrix}
		\epsilon_{S,u}\\
		\epsilon_{P,u}\\
		\epsilon_{T,u}\\
	\end{pmatrix}_{(\mu)} &= [\frac{\alpha_e(\mu)}{2\pi}\begin{pmatrix}
		-4/9 & 0 & 0 \\
		0 & -4/9 & 0\\
		0 &  0 & 5/9\\
	\end{pmatrix}
	+\frac{\alpha_3(\mu)}{2\pi}\begin{pmatrix}
		-4 & 0 & 0 \\
		0 & -4 &  0\\
		0 &  0 & 4/3\\
	\end{pmatrix}] \begin{pmatrix}
		\epsilon_{S,u}\\
		\epsilon_{P,u}\\
		\epsilon_{T,u}\\
	\end{pmatrix}_{(\mu)},
\end{split}
\label{eq:adm_eu}
\eeq
where the flavor indices are implicit. The QED and weak couplings are important as they introduce mixing between different operators. Solving the differential equations with the three-loop $\beta$-functions and taking into account the top and bottom quark mass thresholds, we obtain the numerical relations between effective couplings at different energy scales:
\beq
\begin{pmatrix}
	C_{NLQu}  \\
	C_{NLdQ} \\
	C^\prime_{NLdQ} \\
\end{pmatrix}_{(\mu = M_Z)} = \begin{pmatrix}
	1.18 & 0 & 0 \\
	0 & 1.18 & -0.117\\
	0 &  -2.44\times 10^{-3} & 0.966\\
\end{pmatrix} \begin{pmatrix}
	C_{NLQu}  \\
	C_{NLdQ} \\
	C^\prime_{NLdQ} \\
\end{pmatrix}_{(\mu = \text{1 TeV})},
\eeq
\beq
\begin{pmatrix}
	\epsilon_{S,du}  \\
	\epsilon_{P,du} \\
	\epsilon_{T,du} \\
\end{pmatrix}_{(\mu = \text{2 GeV})} = \begin{pmatrix}
	1.52 & 2.34\times 10^{-6} & -0.0218 \\
	2.34\times 10^{-6} & 1.52 & -0.0218\\
	-2.26\times 10^{-4} & -2.26\times 10^{-4}  & 0.878\\
\end{pmatrix} \begin{pmatrix}
	\epsilon_{S,du}  \\
	\epsilon_{P,du} \\
	\epsilon_{T,du} \\
\end{pmatrix}_{(\mu = M_Z)},
\eeq
\beq
\begin{pmatrix}
	\epsilon_{S,d}  \\
	\epsilon_{P,d} \\
	\epsilon_{T,d} \\
\end{pmatrix}_{(\mu = \text{2 GeV})} = \begin{pmatrix}
	1.52 & 0 & 0 \\
	0 & 1.52 & 0\\
	0 & 0  & 0.869\\
\end{pmatrix} \begin{pmatrix}
	\epsilon_{S,d}  \\
	\epsilon_{P,d} \\
	\epsilon_{T,d} \\
\end{pmatrix}_{(\mu = M_Z)},
\eeq
\beq
\begin{pmatrix}
	\epsilon_{S,u}  \\
	\epsilon_{P,u} \\
	\epsilon_{T,u} \\
\end{pmatrix}_{(\mu = \text{2 GeV})} = \begin{pmatrix}
	1.53 & 0 & 0 \\
	0 & 1.53 & 0\\
	0 & 0  & 0.867\\
\end{pmatrix} \begin{pmatrix}
	\epsilon_{S,u}  \\
	\epsilon_{P,u} \\
	\epsilon_{T,u} \\
\end{pmatrix}_{(\mu = M_Z)}.
\eeq
The numerical relations between LEFT WCs at 2~GeV and SMNEFT WCs at 1~TeV, with $V_{ud} = 0.97420$~\cite{Tanabashi:2018oca}, are
\beq
\begin{split}
	&\epsilon_{S,du} = -1.84 C_{NLQu}+1.79 C_{NLdQ} -0.199 C^\prime_{NLdQ}\,,\\
	&\epsilon_{P,du} = 1.84 C_{NLQu}+1.79 C_{NLdQ} -0.157 C^\prime_{NLdQ}\,,\\
	&\epsilon_{T,du} = 5.49\times 10^{-4} C_{NLQu}-2.14\times 10^{-3} C_{NLdQ} +0.849 C^\prime_{NLdQ}\,,\\
	&\epsilon_{S,u} = -1.76 C_{NLQu}\,,\\
	&\epsilon_{P,u} = 1.76 C_{NLQu}\,,\\
	&\epsilon_{T,u} = 0\,,\\
	&\epsilon_{S,d} = -1.80 C_{NLdQ}+0.179 C^\prime_{NLdQ}\,,\\
	&\epsilon_{P,d} = -1.80 C_{NLdQ}+0.179 C^\prime_{NLdQ}\,,\\
	&\epsilon_{T,d} =  2.12\times 10^{-3} C_{NLdQ}-0.839 C^\prime_{NLdQ}\,.\\
\end{split}
\label{Eq:eps_C}
\eeq

Low energy constraints on the SMNEFT WCs from nuclear beta decay, pseudoscalar meson decay, and coherent scattering have been discussed in Ref.~\cite{Bischer:2019ttk} without accounting for the effects of RG running. The RG running is crucial, as it introduces operator mixing which produces degeneracies in the WCs. Here we first calculate the LEFT and SMNEFT WCs below and above the electroweak scale, respectively. After the RG running, we convert the low energy constraints on the LEFT WCs to the high energy constraints on the SMNEFT WCs, and compare them with those from high energy collider experiments at the same energy scale.

\section{Low-energy constraints}
\label{sec:low}

\subsection{Neutrino mass bounds}
Scalar and tensor interactions that flip the neutrino chirality contribute to the neutrino mass radiatively. Both one- and two-loop corrections to the neutrino mass can be generated by chirality-changing operators. Here we ignore the one-loop corrections since, except for the top quark, they are (counterintuitively) suppressed by a factor of $(m_q/M_Z)^2$ as compared to the two-loop corrections~\cite{Prezeau:2004md,Ito:2004sh}. The two-loop contribution is
estimated as
\beqa
\Delta m_\nu &\simeq& 3g^2 G_F \epsilon \frac{m_q M_W^2}{(4\pi)^4}(\text{ln}\frac{\mu^2}{M_W^2})^2\,,
\eeqa
where $m_q$ is a quark mass, 
$\mu$ is the renormalization scale, and $\epsilon$ can be either a NC or CC GNI parameter. We conservatively take $\mu$ to not
be too far above the electroweak scale so that the top quark loop correction is suppressed.

Bounds from neutrino masses and oscillations are very model specific because of the importance of the properties of the particles in the loops and the possibility of cancellations between loop and other contributions. However, barring fine-tuned cancellations, they provide an order of magnitude estimate of how much the new interactions may contribute to neutrino masses. For our estimates, we assume neutrinos acquire mass only from loop effects due to the new interactions, i.e., neutrino masses vanish as $\epsilon \to 0$.
Then, constraints on the contact interactions can be obtained by requiring $\Delta m_\nu < \sum m_\nu$. A recent upper bound on the sum of neutrino mass from cosmological observations and particle physics experiments is $\sum m_\nu \lesssim 0.26$~eV~\cite{Loureiro:2018pdz}, which is model dependent. The most recent model-independent bound is that obtained by the KATRIN Collaboration~\cite{Aker:2019uuj}. They reported a 1.1~eV upper bound on the effective neutrino mass based on the $\beta$-decay electron spectrum.  
The bounds on the scalar and tensor contact interactions from neutrino masses without (with) cosmological inputs are
\begin{eqnarray}
|\epsilon^{\alpha\beta11}_{S,P,T}| \lesssim  10^{-3}\, (10^{-4})\,,\quad \,|\epsilon^{\alpha\beta22}_{S,P,T}| \lesssim  10^{-5}\, (10^{-6})\,,\quad
 |\epsilon^{\alpha\beta33}_{S,P,T} | \lesssim  10^{-6}\,(10^{-7})\,.
\label{eq:eps23}
\end{eqnarray}
The bounds using cosmological data are only suggestive  because we have not evaluated how the relic neutrino abundance is affected by the new interactions.
From Eq.~(\ref{eq:eps23}), we see that if GNI are also coupled to heavy quark flavors, the bounds on the SMNEFT WCs $C_{NLQu},\; C_{NLdQ},\;\text{and}\, C^\prime_{NLdQ}$ are too strong to be probed by other experiments, current or future.  Despite the highly model-dependent nature of this conclusion, we focus on couplings to first generation quarks in the rest of the paper.

Related bounds arise from neutrino magnetic moments via an external photon attached to the fermion loop responsible for neutrino mass generation. The magnetic moment induced by scalar and tensor GNI is bounded by~\cite{Xu:2019dxe}
\beq
\mu_\nu \approx \frac{e G_F m_d}{8 \pi^2}\epsilon \lesssim 3\times 10^{-11}\mu_B\,,
\eeq
where the Bohr magneton $\mu_B = \frac{e\hbar}{2m_e c} \simeq 2.9 \times 10^{-7}\text{ eV}^{-1}$. This yields
\beq
\mid\epsilon^{\alpha\beta11}_{S,P,T} \mid \lesssim 30\,,
\eeq
which are much weaker than the bounds above.

\subsection{Pseudoscalar meson decay}
\begin{figure}[h!]
	\centering
	\begin{subfigure}{.49\textwidth}
		\centering
		\includegraphics[width=\textwidth]{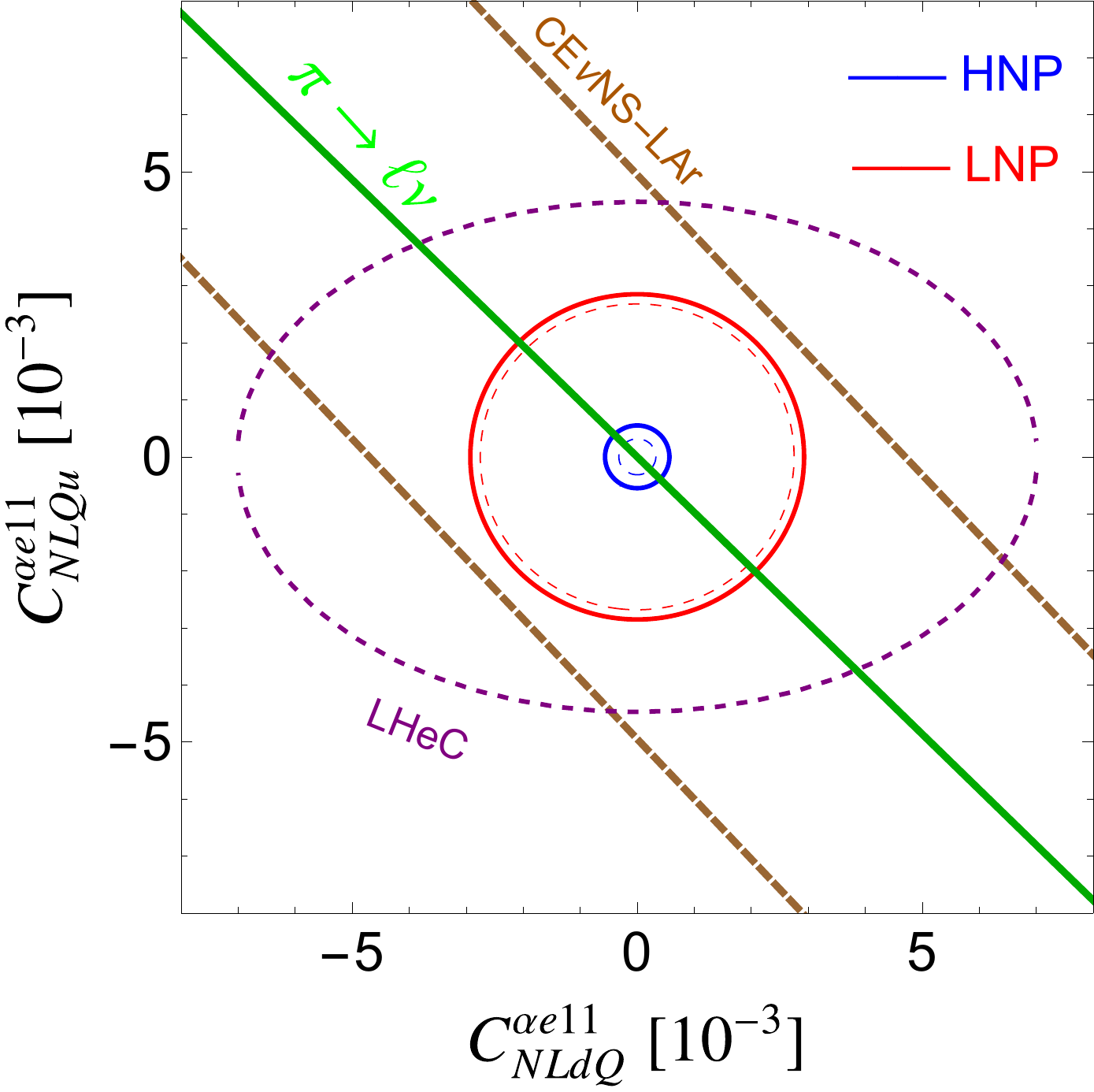}
	\end{subfigure}
	\begin{subfigure}{.49\textwidth}
		\centering
		\includegraphics[width=\textwidth]{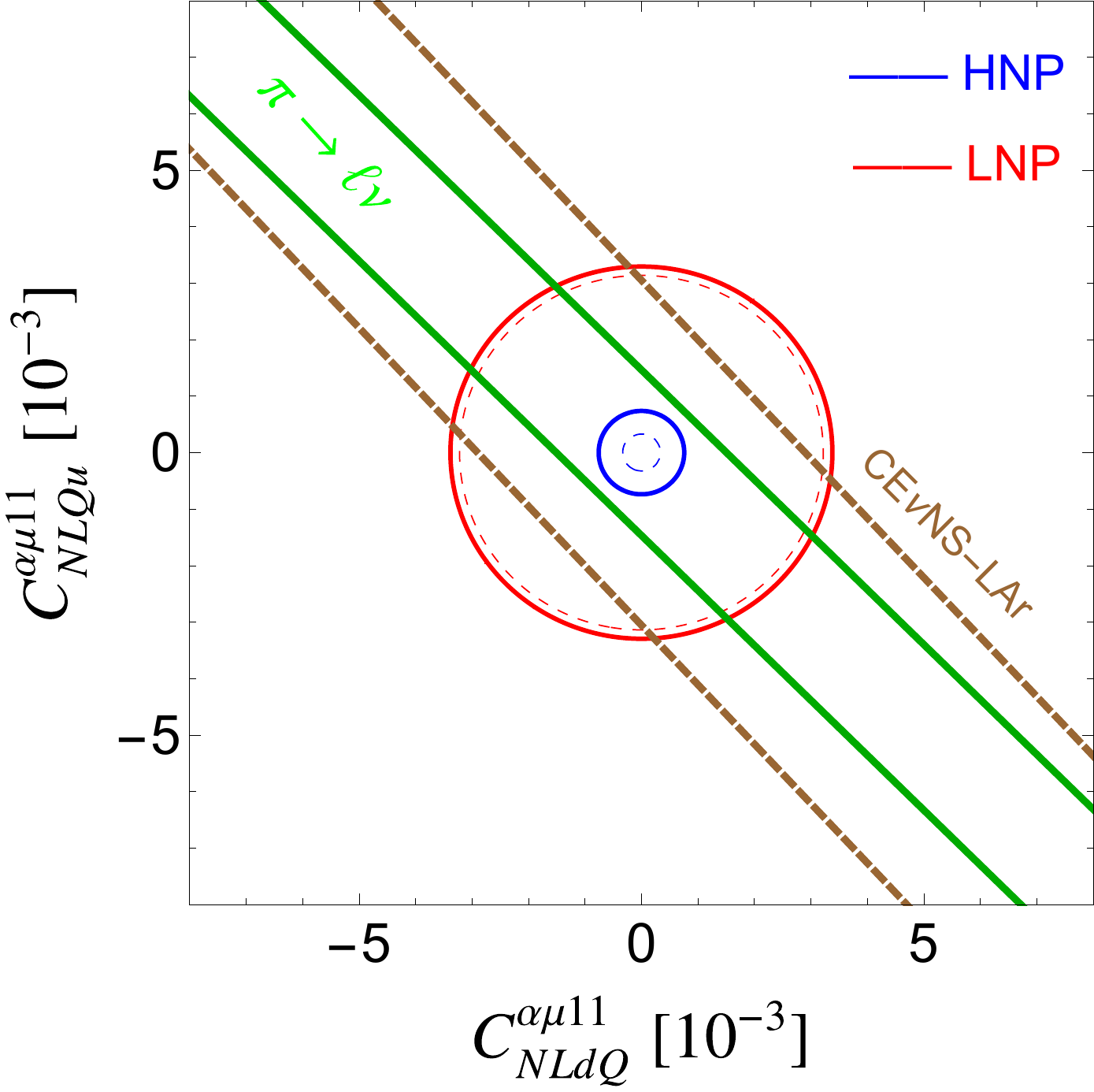}
	\end{subfigure}
	\begin{subfigure}{.49\textwidth}
		\centering
		\includegraphics[width=\textwidth]{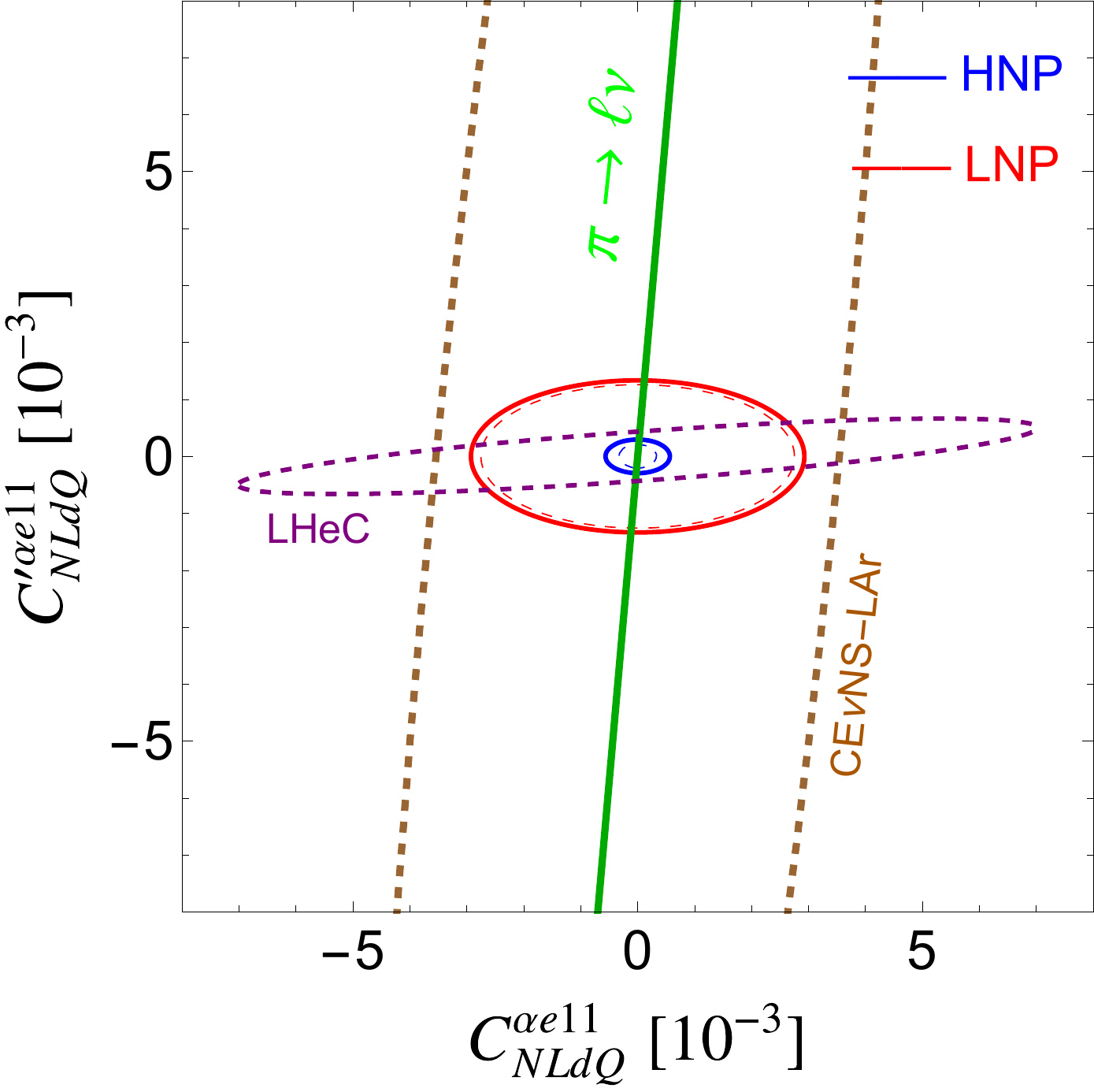}
	\end{subfigure}
	\begin{subfigure}{.49\textwidth}
		\centering
		\includegraphics[width=\textwidth]{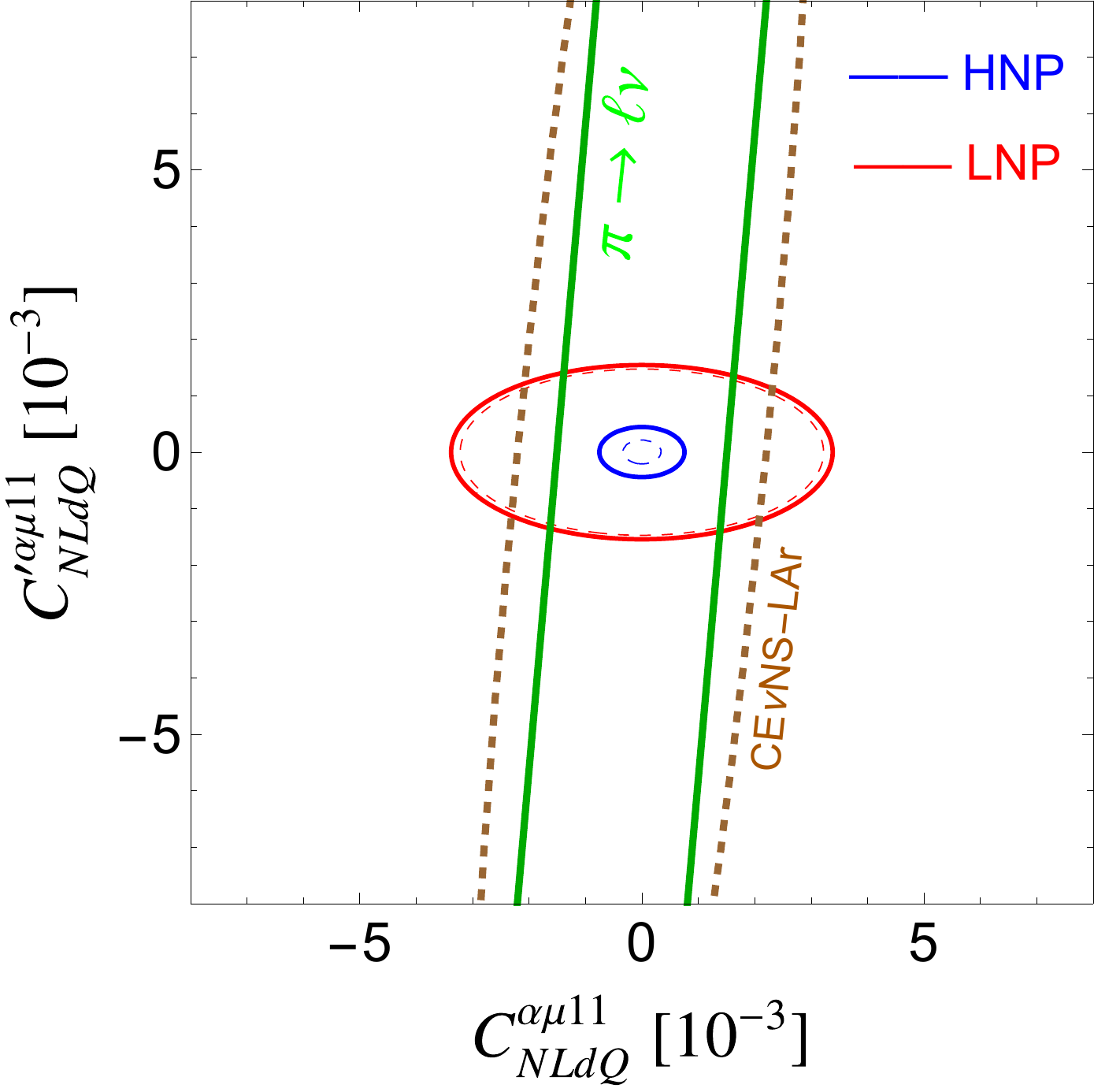}
	\end{subfigure}
	\caption{The 90\%~C.L. allowed regions in the $C_{NLdQ}$-$C_{NLQu}$ planes (upper panels) and $C_{NLdQ}$-$C^\prime_{NLdQ}$ planes (lower panels) at 1~TeV with electron flavor (left panels) and muon flavor (right panels). The green lines (overlapping in the left panels) are the bounds from pion decay with the third parameter  set to zero to break the degeneracy. The red (blue) solid contours correspond to current  LHC  searches with $L = 139\,\text{fb}^{-1}$ for the low-scale new physics LNP (high-scale new physics HNP) case. The brown dashed lines show the projected bounds from the future LAr COHERENT experiment, with $C_{NLQu}$ is set to zero in the lower panels to obtain meaningful bounds. The red and blues dashed contours are the projected bounds from HL-LHC with 3~$\text{ab}^{-1}$ of data.   The dashed purple contours in the left panels are the projected bounds from LHeC with 3~$\text{ab}^{-1}$.}
	\label{fig:Coll_Scalar_EFT}
\end{figure}

The pseudoscalar quark bilinear can contribute to the leptonic decay of a pseudoscalar meson ($P$). In the SM, the decay is helicity suppressed so that the width  $\Gamma_{\text{SM}} (P\rightarrow \ell \nu)\propto m_\ell^2$. The suppression is lifted by pseudoscalar GNI
\beq
 \Gamma_{\text{GNI,p}}(P\rightarrow \ell_\beta \nu_\alpha)\propto (\epsilon_{P,du}^{\alpha\beta11})^2 \frac{m_\pi^4}{(m_u+m_d)^2}\,.
 \eeq
  The branching ratio
\beqa
R_{\pi} \equiv \frac{\Gamma(\pi\rightarrow e\nu[\gamma])}{\Gamma(\pi\rightarrow \mu\nu[\gamma])} = R^{(0)}_{\pi}[1+\Delta_\pi], \quad
{\rm with}\ \  R^{(0)}_{\pi} = \frac{m_e^2}{m_\mu^2}(\frac{m_\pi^2-m_e^2}{m_\pi^2-m_\mu^2})^2\,,
\eeqa 
serves as a good observable, as the experiment systematic uncertainties shared by the two processes cancel in the ratio. $\Delta_\pi$ contains higher order corrections~\cite{Cirigliano:2007xi}. $\Gamma_{(\pi \rightarrow \ell\overline{\nu}[\gamma])}$ contains physical and virtual photons (radiative corrections). Including pseudoscalar GNI interactions~\cite{Cirigliano:2013xha},
\beq
\frac{R_\pi}{R_\pi^{SM}} = \frac{1 + \mid \frac{B_0}{m_e} \epsilon_{P,du}^{\alpha e11} \mid^2 }{1 + \mid \frac{B_0}{m_\mu} \epsilon_{P,du}^{\alpha\mu11} \mid^2}\,,
\label{rat}
\eeq
where $B_0(\mu) = m_\pi^2/(m_u(\mu)+m_d(\mu))$. 
Taking $m_\pi = 139.57$~MeV, $m_u^{\overline{\text{MS}}} (\mu=\text{2 GeV})= 2.16$~MeV and $m_d^{\overline{\text{MS}}} (\mu=\text{2 GeV})= 4.67$~MeV~\cite{Tanabashi:2018oca}, gives $B_0^{\overline{\text{MS}}} (\mu=\text{2 GeV})= 2.8\times 10^3$~MeV. The current combined uncertainty in $R_\pi^{\text{exp}}$~\cite{Britton:1992pg, Britton:1993cj, Czapek:1993kc,Tanabashi:2018oca} and $R_\pi^{\text{SM}}$~\cite{Cirigliano:2007xi, Cirigliano:2007ga} are
\beq
R_\pi = 1.2327(23)\times 10^{-4}\,,\quad  R_\pi^{SM} = 1.2352(1)\times 10^{-4}\,. 
\eeq
If both $\epsilon^{\alpha e11}_{P,du}$ and $\epsilon^{\alpha\mu11}_{P,du}$ are allowed to vary simultaneously, no bound on either parameter
is obtained because they are degenerate, as is evident from Eq.~(\ref{rat}).  With the assumption that only one of $\epsilon_{P,du}$ is nonzero, the 90\% C.L. bounds are
\beq
\mid \epsilon_{P,du}^{\alpha e11}\mid < 6.2 \times 10^{-6}, \quad \text{and} \quad \mid \epsilon_{P,du}^{\alpha\mu11}\mid < 2.7  \times 10^{-3}\,.
\eeq
Because the measured branching to the electron channel is tiny, $\epsilon_{P,du}^{\alpha e11}$ is highly constrained.
These bounds are much stronger than the ones obtained in Ref.~\cite{Cirigliano:2013xha}, which assumed that both $\epsilon_{P,du}$ and 
$\tilde{\epsilon}_{P,du}$ are simultaneously nonzero, which however, cannot be realized with the three SMNEFT operators considered here. The bounds on the coefficients of the low-energy effective Lagrangian can be translated to bounds on the three SMNEFT WCs by adopting the relations in Eq.~(\ref{Eq:eps_C}), which display degeneracies between the SMNEFT WCs. We therefore bound the individual WCs by setting the other two to zero. 
The 90\% C.L. bounds on the SMNEFT WCs are
\beq
\mid C^{\alpha e11}_{NLQu} \mid < 3.3 \times 10^{-6}\,,\quad\mid C^{\alpha e11}_{NLdQ} \mid < 3.4 \times 10^{-6}\,,\quad \mid C^{\prime \alpha e11}_{NLdQ} \mid < 3.9 \times 10^{-5}\,,
\eeq
\beq
 \mid C^{\alpha\mu11}_{NLQu} \mid < 1.5 \times 10^{-3}\,,\quad \mid C^{\alpha\mu11}_{NLdQ} \mid < 1.5  \times 10^{-3}\,,\quad \mid C^{\prime\alpha\mu11}_{NLQu} \mid < 1.7 \times 10^{-2}\,.
\eeq
The correlations between the $C_{NLdQ}$ and $C_{NLQu}$ ($C^\prime_{NLdQ}$), with $C^\prime_{NLdQ}$ ($C_{NLQu}$) set to zero,  are shown by the green lines in the upper (lower) panel of Fig.~\ref{fig:Coll_Scalar_EFT}. 

To circumvent the degeneracy in $\epsilon^{\alpha e11}_{P,du}$ and $\epsilon^{\alpha\mu11}_{P,du}$ in Eq.~(\ref{rat}), we now apply the individual decay width measurements of $\pi \rightarrow \ell \nu[\gamma]$ to set the bounds. 
	In the SM, the decay width at tree level is 
	\beq
	\Gamma_{(\pi \rightarrow \ell\overline{\nu}[\gamma])} = \frac{G_F^2}{8\pi}f_\pi^2V_{ud}^2m_\ell^2m_\pi(1-\frac{m_\ell^2}{m_\pi^2})^2(1+\Delta_\pi)\,.
	\eeq
	The theoretical uncertainties are mainly from calculations of the decay constant and radiative corrections. According to Refs.~\cite{Tanabashi:2018oca,Ananthanarayan:2004qk,DescotesGenon:2005pw,Cirigliano:2007ga,Aoki:2019cca},
	\beq
	\Delta_\pi = 0.0176\pm 0.0021\,,\quad  f_{\pi}= 130.2  \pm 1.2\text{ MeV}\,.
	\eeq 
	The universal theoretical uncertainties yield
	\beq
	\frac{\delta \Gamma_{(\pi^+\rightarrow e^+\nu_e [\gamma])}}{\Gamma_{(\pi^+\rightarrow e^+\nu_e [\gamma])}} =\frac{\delta \Gamma_{(\pi^+\rightarrow \mu^+\nu_\mu [\gamma])}}{\Gamma_{(\pi^+\rightarrow \mu^+\nu_\mu[\gamma])}} = 1.9\times 10^{-2}\,.
	\eeq 	
	Measurements give $BR(\pi^+ \rightarrow e^+ \nu_e[\gamma]) = (1.230 \pm 0.004)\times 10^{-4}\%$, $BR(\pi^+ \rightarrow \mu^+ \nu_\mu[\gamma]) = (99.98770 \pm 0.00004)\%$, and $\tau_{\pi^\pm} = 26.033(5)$ ns~\cite{Tanabashi:2018oca}. The experimental uncertainties in the electron (muon) channel is $3.3\times 10^{-3}$ ($1.9\times10^{-4}$) and can be neglected.
Assuming that the new physics contributions do not exceed the theoretical uncertainties, the bounds on $\epsilon_{P,du}^{\alpha e11}$ and $\epsilon_{P,du}^{\alpha\mu11}$ are given by
	\beq
	\mid\frac{B_0}{m_e}\epsilon_{P,du}^{\alpha e11}\mid^2<\frac{\delta \Gamma_{(\pi^+\rightarrow e^+\nu_e [\gamma])}}{\Gamma_{(\pi^+\rightarrow e^+\nu_e [\gamma])}} ,\qquad 
	\mid\frac{B_0}{m_\mu}\epsilon_{P,du}^{\alpha\mu11}\mid^2 < \frac{\delta \Gamma_{(\pi^+\rightarrow \mu^+\nu_\mu [\gamma])}}{\Gamma_{(\pi^+\rightarrow \mu^+\nu_\mu [\gamma])}} \,,
	\eeq
	which yield
	\beq
	\mid \epsilon_{P,du}^{\alpha e11}\mid < 3.4 \times 10^{-5}\,,\quad \mid \epsilon_{P,du}^{\alpha\mu11}\mid < 6.6 \times 10^{-3}\,,
	\eeq
	at the 90\%~C.L.
	By allowing only one WC to be nonzero at a time, the 90\%~C.L. bounds on the SMNEFT WCs derived from individual decay channels are
	\beq
	\mid C^{\alpha e11}_{NLQu} \mid < 1.9 \times 10^{-5}\,, \quad \mid C^{\alpha e11}_{NLdQ} \mid < 1.9 \times 10^{-5}\,,\quad \mid C^{\prime \alpha e11}_{NLdQ} \mid < 2.2 \times 10^{-4}\,,
	\eeq
	\beq
	\mid C^{\alpha\mu11}_{NLQu} \mid < 3.6 \times 10^{-3}\,,\quad \mid C^{\alpha\mu11}_{NLdQ} \mid < 3.7  \times 10^{-3}\,,\quad \mid C^{\prime\alpha\mu11}_{NLQu} \mid < 4.2 \times 10^{-2}\,.
	\eeq


\subsection{Nuclear beta decay}
Nuclear $\beta$-decay is another low-energy probe that is sensitive to the new CC GNI interactions. The nucleon-level effective Lagrangian contributing to neutron beta decay, $n\rightarrow p + e^{-}+\overline{\nu}$, is using Eq.~(\ref{Eq:hermitian}),
\beqa
\nonumber
\lag_N = &-&\frac{G_F}{\sqrt{2}} V_{ud} [ \overline{p}\gamma^\mu (g_V - g_A\gamma^5)n\cdot \overline{e}\gamma_\mu(1-\gamma^5)\nu_e
+g_S(\epsilon^{\alpha e11}_{S,du})^*\, \overline{p}n\cdot \overline{e}(1+\gamma^5)\nu_\alpha \\
&-&g_P(\epsilon^{\alpha e11}_{P,du})^*\, \overline{p}\gamma^5n\cdot \overline{e}(1+\gamma^5)\nu_\alpha
+2g_T(\epsilon^{\alpha e11}_{T,du})^*\, \overline{p}\sigma^{\mu\nu}n\cdot \overline{e}\sigma_{\mu\nu}(1+\gamma^5)\nu_\alpha]+ \text{h.c.}\,,
\eeqa
where $g_{V(A)}$ is the (axial-)vector charge and $g_{S,P,T}$ are the nonstandard charges. Neglecting nucleon recoil and the pseudoscalar
contribution in the $q^2\rightarrow 0$ limit, the neutron $\beta$ decay width is
\beq
\Gamma=\frac{G_F^2 V^2_{ud}}{2\pi^3}[g_V^2 (3\lambda^2+1) + g_S^2\mid\epsilon^{\alpha e11}_{S,du}\mid^2 + 48g_T^2\mid\epsilon^{\alpha e11}_{T,du}\mid^2]\, I\,,
\eeq
where $\lambda\equiv g_V/g_A$ and
\beq
I = \int p_e E_e (M_n - M_p - E_e)^2 dE_e \approx 0.06\,\text{MeV}^5\,.
\eeq
The decay width can also be written in terms of the NC effective couplings by using the relations in Eq.~(\ref{eq:NCCC}):
\beq
\Gamma=\frac{G_F^2 V^2_{ud}}{2\pi^3}[g_V^2 (3\lambda^2+1) + g_S^2(\frac{\epsilon^{\alpha e11}_{S,u}}{V_{ud}^2}-\epsilon^{\alpha e11}_{S,d})^2 + 48g_T^2(\epsilon^{\alpha e11}_{T,d})^2]\, I\,.
\eeq
From Ref.~\cite{Gonzalez-Alonso:2018omy}, the 90\% C.L. bounds, based on the differential observables from polarized nuclear beta decay, are
\beq
\mid\epsilon_{S,du}^{\alpha e11}\mid < 0.063\,, \quad \mid\epsilon_{T,du}^{\alpha e11}\mid < 0.024\,.
\label{bouns:eps_STdu_e}
\eeq
Bounds on the NC parameters can be computed by using the relations in Eq.~(\ref{eq:NCCC}) with $\epsilon^{\alpha e11}_{P,du}$ taken to be 0:
\beq
\mid\epsilon_{S,d}^{\alpha e11}\mid\,,\mid \epsilon_{P,d}^{\alpha e11}\mid < 0.063\,,\quad \mid\epsilon_{S,u}^{\alpha e11}\mid\,, \mid\epsilon_{P,u}^{\alpha e11}\mid < 0.060\,,\quad \mid\epsilon_{T,d}^{\alpha e11}\mid < 0.024\,.
\eeq
Degeneracies do not permit simultaneous bounds on all the SMNEFT WCs.   With the assumption that only one of them is nonzero, the 
90\%~C.L. bounds are
\beq
\mid C^{\alpha e11}_{NLQu} \mid < 3.4 \times 10^{-2}\,, \quad  \mid C^{\alpha e11}_{NLdQ} \mid < 3.5 \times 10^{-2}\,, \quad \mid C^{\prime \alpha e11}_{NLdQ} \mid < 2.8 \times 10^{-2}\,.
\eeq
These constraints are much weaker than the ones from charged pion decay. 

\subsection{Neutrino deep inelastic scattering}
Neutrino deep inelastic scattering on nucleons can be modified by scalar, pseudoscalar, and tensor GNI. Note that the charged current is not affected by the three GNI considered as the right-handed neutrino is absent in the neutrino beams. The total charged current and neutral current neutrino-nucleon scattering cross sections in the SM are
\beq
\sigma^{CC}_{\nu N, SM} = \frac{ 2 G_F^2}{\pi}E_\nu M_N[\langle x d_N + x\frac{1}{3}\overline{u}_N\rangle],
\label{eq:CC1}
\eeq 
\beq
\sigma^{CC}_{\overline{\nu} N, SM} = \frac{2 G_F^2}{\pi}E_\nu M_N[\langle x \frac{1}{3}u_N + x\overline{d}_N\rangle],
\label{eq:CC2}
\eeq 
\beq
\begin{split}
\sigma^{NC}_{\nu N, SM} = \frac{2 G_F^2}{\pi}E_\nu M_N [&(g^2_{L,u} + \frac{1}{3}g^2_{R,u})\langle x u_N \rangle + (g^2_{L,d} + \frac{1}{3}g^2_{R,d})\langle x d_N \rangle \\
+ &(g^2_{R,u} + \frac{1}{3}g^2_{L,u})\langle x \overline{u}_N \rangle+(g^2_{R,d} + \frac{1}{3}g^2_{L,d}) \langle x \overline{d}_N\rangle],
\end{split}
\label{eq:NC1}
\eeq 
\beq
\begin{split}
	\sigma^{NC}_{\overline{\nu} N, SM} = \frac{2 G_F^2}{\pi}E_\nu M_N [&(g^2_{R,u} + \frac{1}{3}g^2_{L,u})\langle x u_N \rangle + (g^2_{R,d} + \frac{1}{3}g^2_{L,d})\langle x d_N \rangle \\
	+ &(g^2_{L,u} + \frac{1}{3}g^2_{R,u})\langle x \overline{u}_N \rangle+(g^2_{L,d} + \frac{1}{3}g^2_{R,d})\langle x \overline{d}_N\rangle],
\end{split}
\label{eq:NC2}
\eeq 
where we have neglected contributions from heavy quarks, and 
\beq
\langle x q_N\rangle \equiv \int_{0}^{1} x q_N(x) dx,\quad \langle x \overline{q}_N\rangle \equiv \int_{0}^{1} x \overline{q}_N(x) dx\,,
\eeq
determine the fraction of nucleon momentum carried by quarks and anti-quarks.
$g_{L,f}$ and $g_{R,f}$ are the SM effective couplings given in Eq.~(\ref{eq:eff_coup}). We take~\cite{Erler:2013xha}
\beq
g_{L,u} = 0.3457\,,\quad g_{R,u} = -0.1553\,,\quad g_{L,d} = -0.4288\,,\quad g_{R,d} = 0.0777\,,
\label{eq:geff}
\eeq
which include the one-loop and leading two-loop corrections.  The neutral current is modified by scalar, pseudoscalar, and tensor GNI: 
\beqa
\label{eq:NC3}
&& \sigma^{NC}_{\nu N, S(P)} = \sigma^{NC}_{\overline{\nu} N, S(P)} = \frac{ G_F^2}{12 \pi}E_\nu M_N [\epsilon_{S(P),u}^2\langle x (u_N + \overline{u}_N) \rangle +\epsilon_{S(P),d}^2\langle x (d_N + \overline{d}_N) \rangle]\,, \\
&& \sigma^{NC}_{\nu N, T} = \sigma^{NC}_{\overline{\nu} N, T} = \frac{56 G_F^2}{3 \pi}E_\nu M_N [\epsilon_{T,u}^2\langle x (u_N + \overline{u}_N) \rangle +\epsilon_{T,d}^2\langle x (d_N + \overline{d}_N) \rangle]\,,
\label{eq:NC4}
\eeqa
where the flavor indices are suppressed for simplicity. In the following analysis, we assume the target is isoscalar and composed of free nucleons, so that we may use the proton PDF. Under these assumptions, the nuclear PDFs  become
\beq
\langle x d_N \rangle = \langle x u_N \rangle  = \frac{N}{2} \langle x (u_p+ d_p) \rangle\,,\quad \langle x \overline{d}_N \rangle = \langle x \overline{u}_N \rangle  = \frac{N}{2} \langle x (\overline{d}_p+ \overline{u}_p) \rangle\,.
\eeq

\subsubsection{CHARM: $\nu_e q \rightarrow \nu q$}
The CHARM collaboration measured the ratio of total cross sections for semileptonic $\nu_e$ and $\overline{\nu}_e$ scattering to be~\cite{Dorenbosch:1986tb}
\beq
R^e \equiv \frac{\sigma(\nu_e N \rightarrow \nu X) + \sigma(\overline{\nu}_e N \rightarrow \nu X)}{\sigma(\nu_e N \rightarrow e^- X)+\sigma(\overline{\nu}_e N \rightarrow e^+ X)}=0.406 \pm 0.140\,.
\eeq
The SM prediction from Eqs.~(\ref{eq:CC1}) to (\ref{eq:NC2}) is 
\beq
R^e = g_L^2 + g_R^2 = 0.3335\,,
\eeq
where
\beq
g_L^2 = g_{L,u}^2 + g_{L,d}^2,\quad  g_R^2 = g_{R,u}^2 + g_{R,d}^2\,.
\eeq
Including the new GNI contributions from Eqs.~(\ref{eq:NC3}) and (\ref{eq:NC4}), $R^e$ becomes
\beq
R^e = g_L^2 + g_R^2 + \frac{1}{12}\sum\limits_{q=u,d} ((\epsilon^{\alpha e11}_{S,q})^2+(\epsilon^{\alpha e11}_{P,q})^2+224(\epsilon^{\alpha e11}_{T,q})^2)\,.
\eeq
The 90\% C.L. bounds on the LEFT parameters are
\beq
\mid\epsilon^{\alpha e11}_{S,q}\mid\,,\mid\epsilon^{\alpha e11}_{P,q}\mid < 1.9\,,\quad \mid\epsilon^{\alpha e11}_{T,q}\mid < 0.13\,.
\eeq
With only a single constraint on $R^e$, the degeneracy between the three SMNEFT WCs remains unbroken. The bounds on the SMNEFT WCs, with the assumption that only one of the WCs is nonzero at a time, are	
\beq
\mid C^{\alpha e11}_{NLQu}\mid < 0.77\,,\quad \mid C^{\alpha e11}_{NLdQ}\mid  < 0.75\,,\quad \mid C^{\prime \alpha e11}_{NLdQ}\mid < 0.15\,,
\eeq
which are much weaker than the bounds from charged pion decay and nuclear beta decay.

\subsubsection{NuTeV: $\nu_\mu q \rightarrow \nu q$}
The NuTeV collaboration has measured the ratios of neutral current to charged current neutrino-nucleon cross sections~\cite{Zeller:2001hh}:
\beqa
\label{eq:Rnu}
R^{\nu}\equiv \frac{\sigma(\nu_\mu N \rightarrow \nu X)}{\sigma(\nu N \rightarrow \mu^- X)}=0.3916 \pm 0.0013 , \quad 
%
R^{\overline{\nu}}\equiv \frac{\sigma(\overline{\nu}_\mu N \rightarrow \overline{\nu} X)}{\sigma(\overline{\nu} N \rightarrow \mu^+ X)}= 0.4050 \pm 0.0027\,.~~~
\eeqa
In the SM, the cross section ratios on an isoscalar target composed of free nucleons are
\beq
	R_{SM}^{\nu} = \frac{(g_{L}^2 + \frac{1}{3} g_{R}^2)f_q+(g_{R}^2 + \frac{1}{3} g_{L}^2)f_{\overline{q}}}{f_q+\frac{1}{3}f_{\overline{q}}}, \quad 
%
R_{SM}^{\overline{\nu}} = \frac{(g_{R}^2 + \frac{1}{3} g_{L}^2)f_q+(g_{L}^2 + \frac{1}{3} g_{R}^2)f_{\overline{q}} }{\frac{1}{3} f_q+f_{\overline{q}}}\,,\\
\eeq
where $f_q$ and $f_{\overline{q}}$ determine the fraction of proton momentum carried by the first generation of quarks and anti-quarks:
\beq
f_q = \langle x u + x d \rangle = 0.42,\,f_{\overline{q}}=\langle x\overline{u} + x\overline{d} \rangle=0.068\,.
\label{eq:fq}
\eeq
Here we used the CT10 PDFs~\cite{Lai:2010vv} and the Mathematica package ManeParse~\cite{Clark:2016jgm} to obtain the numerical values of $f_q$ and $f_{\overline{q}}$ at $Q^2 = 20\,\text{GeV}^2$. After including the contributions from scalar, pseudoscalar, and tensor GNI, $R^\nu$ and $R^{\overline{\nu}}$ are
\beqa
\label{Eq:Rnu}
R^\nu = \frac{(g_L^2 + \frac{1}{3}g_R^2) f_q+(\frac{1}{3}g_L^2 + g_R^2) f_{\bar{q}} + \frac{1}{24}\sum\limits_{q=u,d}((\epsilon^{\alpha\mu11}_{S,q})^2+(\epsilon^{\alpha\mu11}_{P,q})^2+224(\epsilon^{\alpha\mu11}_{T,q})^2)(f_q+f_{\bar{q}})}{f_q+\frac{1}{3} f_{\bar{q}}},~~~~~\\ 
R^{\overline{\nu}} = \frac{(\frac{1}{3}g_L^2 + g_R^2) f_q+(g_L^2 + \frac{1}{3}g_R^2) f_{\bar{q}} + \frac{1}{24}\sum\limits_{q=u,d}((\epsilon^{\alpha\mu11}_{S,q})^2+(\epsilon^{\alpha\mu11}_{P,q})^2+224(\epsilon^{\alpha\mu11}_{T,q})^2)(f_q+f_{\bar{q}})}{\frac{1}{3}f_q+ f_{\bar{q}}}\,.~~~~~
\label{Eq:Rnubar}
\eeqa
Using the numerical values in Eq.~(\ref{eq:geff}) and~(\ref{eq:fq}), we obtain our naive SM values $R^\nu_{SM} = 0.32$ and $R^{\bar{\nu}}_{SM} = 0.37$, which deviate significantly from the NuTeV measured values in Eq.~(\ref{eq:Rnu}).
Including nuclear effects, partonic charge symmetry violation and strange quarks resolves the NuTeV anomaly~\cite{Bentz:2009yy}, bringing the experimental measurements in good agreement with the SM values $R^\nu = 0.3950$ and $R^{\overline{\nu}} = 0.4066$. We simply rescale our naive SM calculations to the more accurate ones. We apply the same rescaling to the new physics contributions to set the 90\%~C.L. bounds,
\beq
\mid\epsilon^{\alpha\mu11}_{S,q}\mid\,,\mid\epsilon^{\alpha\mu11}_{P,q}\mid < 0.19\,,\quad \mid\epsilon^{\alpha\mu11}_{T,q}\mid < 0.013\,.
\eeq
The degeneracies between the three SMNEFT WCs can be broken by the $R^\nu$ and $R^{\overline{\nu}}$ measurements. By plugging the numerical relations in Eq.~(\ref{Eq:eps_C}) into Eqs.~(\ref{Eq:Rnu}) and (\ref{Eq:Rnubar}), the bounds on the three SMNEFT WCs, allowing all of them to be nonzero simultaneously, are
\beq
\mid C^{\alpha\mu11}_{NLQu}\mid < 0.078\,, \quad \mid C^{\alpha\mu11}_{NLdQ}\mid  < 0.076\,, \quad \mid C^{\prime \alpha\mu11}_{NLdQ}\mid < 0.015\,.
\eeq

\subsection{CE$\nu$NS}
Coherent elastic neutrino-nucleus scattering occurs when the momentum exchanged is smaller than the inverse of the nucleus size, which typically requires neutrino energies of ${\cal{O}}$(10~MeV). The cross section is enhanced by the square of the number of of nucleons, thus providing an excellent tool to investigate GNI at low energies.
The COHERENT experiment has recently observed CE$\nu$NS in a low-threshold CsI detector at the 6.7$\sigma$ level. This is consistent with the SM at 1$\sigma$~\cite{Akimov:2017ade}. The neutrino flux from the Spallation Neutron Source (SNS) is comprised of prompt, monoenergetic $\nu_\mu$ from stopped pion decays, $\pi^+\to \mu^++\nu_\mu$, and $\overline{\nu}_\mu$ and $\nu_e$ from the subsequent muon decays, $\mu^+\to e^++\overline{\nu}_\mu+\nu_e$. 

The neutrino fluxes are 
\begin{align}
\label{eq:nu-spectra.COHERENT}
\phi_{\nu_\mu}(E_{\nu_\mu})&= \mathcal{N} 
\frac{2m_\pi}{m_\pi^2-m_\mu^2}\,
\delta\left(
1-\frac{2E_{\nu_\mu}m_\pi}{m_\pi^2-m_\mu^2}
\right) \ ,
\nonumber\\
\phi_{\nu_e}(E_{\nu_e})&= \mathcal{N} \frac{192}{m_\mu}
\left(\frac{E_{\nu_e}}{m_\mu}\right)^2
\left(\frac{1}{2}-\frac{E_{\nu_e}}{m_\mu}\right)\ ,\\
\phi_{\overline\nu_\mu}(E_{\overline\nu_\mu})&= \mathcal{N}  \frac{64}{m_\mu}
\left(\frac{E_{\overline\nu_\mu}}{m_\mu}\right)^2
\left(\frac{3}{4}-\frac{E_{\overline\nu_\mu}}{m_\mu}\right)\,,\nonumber
\end{align}
where $\mathcal{N}$ is a normalization factor determined by the experimental setup. 
   The $\nu_\mu$ energy is fixed at $(m^2_{\pi}-m_\mu^2)/(2 m_\pi)\approx 30$~MeV due to the two-body pion decay. 
   The $\nu_e$ and $\overline\nu_\mu$ energies have a kinematic upper bound, $m_\mu/2 \approx 50$~MeV. 

The differential cross section including scalar, vector, and tensor contributions reads~\cite{Lindner:2016wff}
\beq
\frac{d\sigma_a^\beta}{dE_r} = \frac{G_F^2}{4\pi}M_{a} N_a^2 [(\xi_S^\beta)^2\frac{E_r}{E_{r,\text{max}}} + (\xi_V^\beta)^2(1-\frac{E_r}{E_{r,\text{max}}}-\frac{E_r}{E_\nu}) + (\xi_T^\beta)^2(1-\frac{E_r}{2 E_{r,\text{max}}}-\frac{E_r}{E_\nu})  ]F^2(q^2)\,,
\eeq
where $a$ denotes the target material and $\alpha$ denotes the neutrino flavor. $M_a$ and $N_a$ are the molar mass of the target nucleus and neutron number of the target, respectively. The flavor index $\beta=\mu$ includes both $\nu_\mu$ and $\overline{\nu}_\mu$. $F(q^2)$ is the nuclear form factor~\cite{Klein:1999gv}. The maximum recoil energy $E_{r,\text{max}}= \frac{2 E_\nu^2}{M_a + 2E_\nu} \approx \frac{2 E_\nu^2}{M_a}$. Since the typical recoil energy $E_r$ is $\cal{O}$(10)~keV, and the neutrino energy $E_\nu$ is $\cal{O}$(10)~MeV, we can safely ignore the interference term between scalar and tensor interactions, which is proportional to $E_r/E_\nu$.
The $\xi_S$, $\xi_V$, and $\xi_T$ collect the contributions from scalar, vector, and tensor interactions, respectively, and are defined as 
\begin{eqnarray}
	(\xi_S^\beta)^2&=&\frac{1}{N_a^2}\{(\sum_{q = u,d} 2 \text{Re}(\epsilon^{\alpha\beta11}_{S,q})[N \frac{m_n}{m_q}f_{Tq}^n + Z \frac{m_p}{m_q} f_{Tq}^p])^2+(\sum_{q = u,d} 2\text{Im}(\epsilon^{\alpha\beta11}_{S,q})[N \frac{m_n}{m_q}f_{Tq}^n + Z \frac{m_p}{m_q} f_{Tq}^p])^2\}\,,\nonumber\\
	(\xi_V^\beta)^2&=&\frac{2}{N_a^2}(Z (2 g_{V,u} + g_{V,d})  + N (g_{V,u} + 2 g_{V,d}) )^2\,,\\
	(\xi_T^\beta)^2&=& \frac{8}{N_a^2} (\sum_{q=u,d} 4\; \text{Re}(\epsilon^{\alpha\beta11}_{T,q})[Z \delta_q^p  + N \delta_q^n])^2\,,\nonumber
	\label{Eq:xi}
\end{eqnarray}
where $f_{Tq}^p$ and $f_{Tq}^n$ are the mass fractions of quark $q$ in the respective nucleon, and the $\delta_q$'s are the corresponding nucleon tensor charges.
The effective vector coupling $g_{V,q}$ is
\beq
g_{V,q}\equiv g_{L,q} + g_{R,q}\,.
\eeq
The expected number of events per day with recoil energy in the energy range 
[$E_r$, $E_r + \Delta E_r$] and arrival time in the time interval [$t$, $t+\Delta t$] is given by
\begin{equation}
\label{eq:recoil-spectrum}
N_{th} (t,E_r,\epsilon)=\sum_{\beta = e,\mu}\frac{m_\text{det}N_A}{M_a}\int_{\Delta E_r}
\,dE_r\int_{\Delta t}\,dt \rho_\alpha(t) \int_{E_\nu^\text{min}}^{E_\nu^\text{max}}\,dE_\nu\,\phi_\beta(E_\nu)
\,\frac{d\sigma^{\beta}_a(\epsilon)}{dE_r}\, ,
\end{equation}
where $m_\text{det}$ is the detector mass, $N_A=6.022\times 10^{23}\,\text{mol}^{-1}$, and $\rho_\alpha(t)$ is the arrival time probability density function. To calculate the differential neutrino-nucleus scattering cross section, we need to evaluate the matrix elements of the operators between nuclear states.
We adopt the following numerical values of the nuclear matrix elements~\cite{Hoferichter:2015dsa,Bhattacharya:2016zcn} 
\beq
\begin{split}
	&f^p_{Tu} = 0.0208\,, \quad f^p_{Td} = 0.0411\,, \quad f^n_{Tu} = 0.0189 \,, \quad f^n_{Td} = 0.0451\,, \\
	&\delta^p_{u} = 0.792\,, \quad \delta^p_{d} = -0.194\,, \quad \delta^n_{u} = -0.194\,,\quad  \delta^n_{d} = 0.792\,.
\end{split}
\eeq

\begin{figure}[tb]
	\centering
	\begin{subfigure}{.49\textwidth}
		\includegraphics[width=\textwidth]{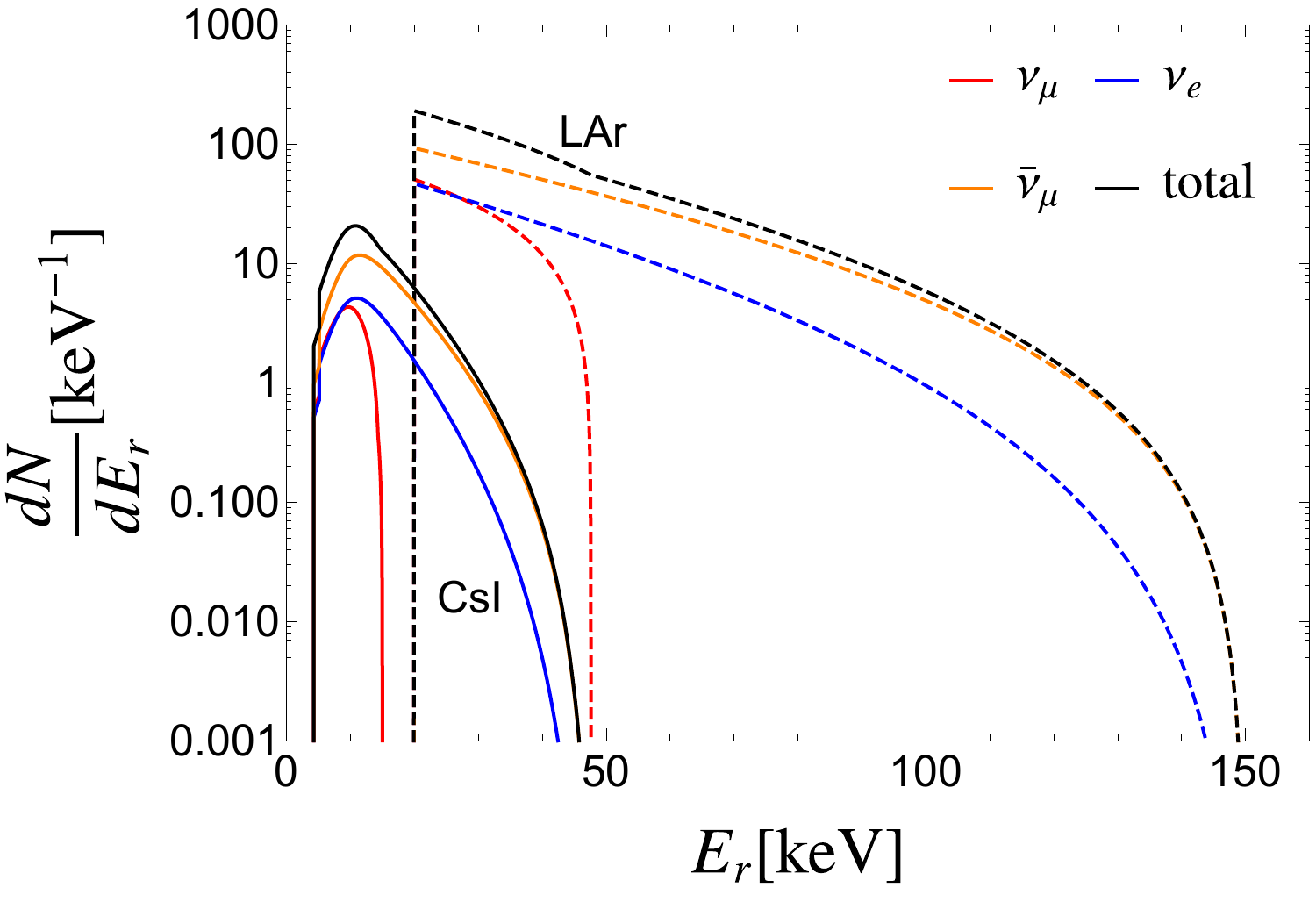}
	\end{subfigure}
	\begin{subfigure}{.49\textwidth}
		\includegraphics[width=\textwidth]{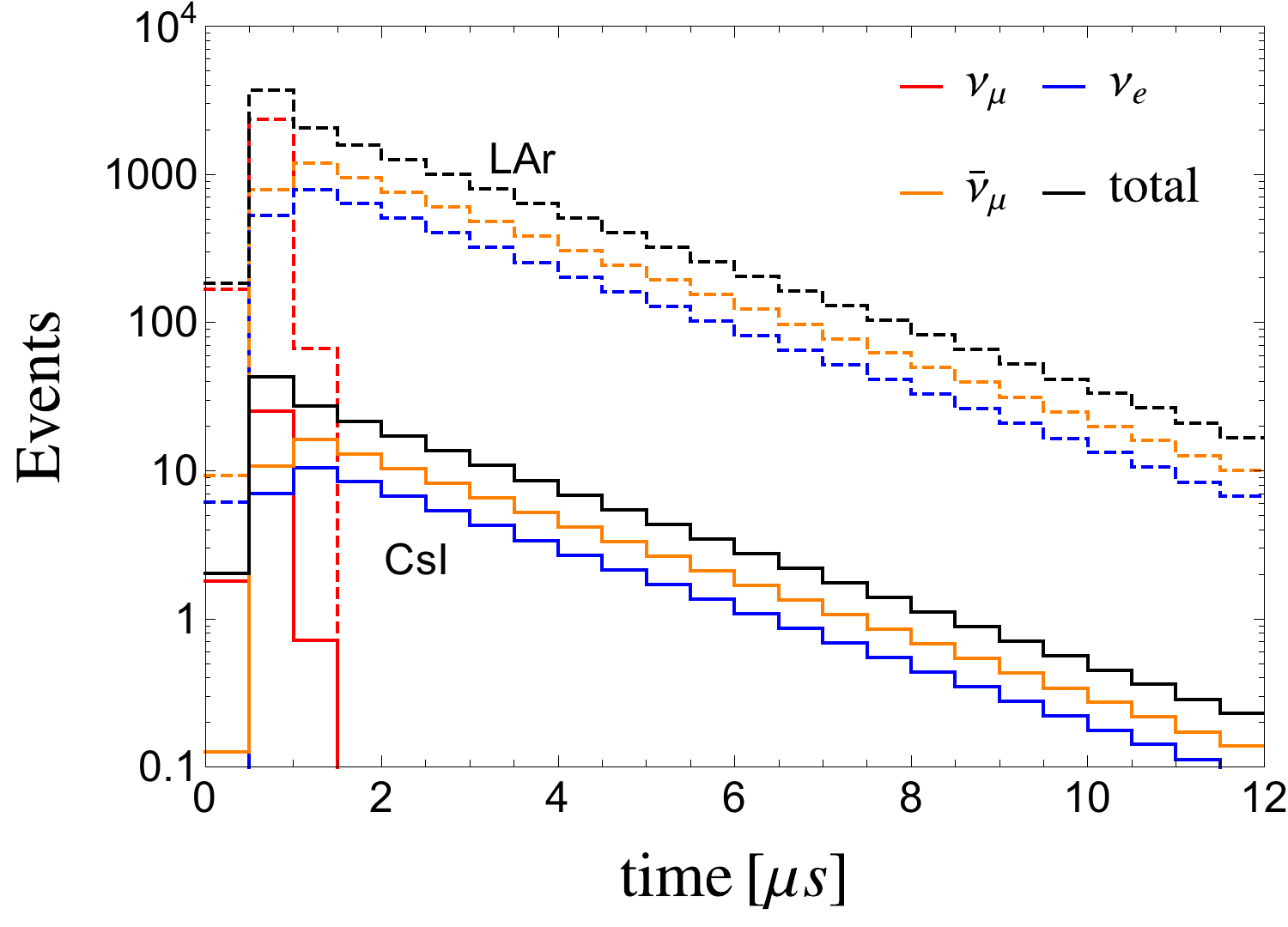}
	\end{subfigure}
	\caption{The SM recoil energy (left) and temporal (right) distributions in the current COHERENT CsI detector (solid lines) and a future COHERENT LAr detector (dashed curves). Threshold effects are included. The red (blue) [orange] curves correspond to the contribution from muon (electron) [anti-muon] neutrinos. The black lines correspond to the sum of all the flavor contributions.}
	\label{fig:CEvNS_dis}
\end{figure} 

\begin{figure}[tb]
	\centering
	\begin{subfigure}{.49\textwidth}
		\includegraphics[width=\textwidth]{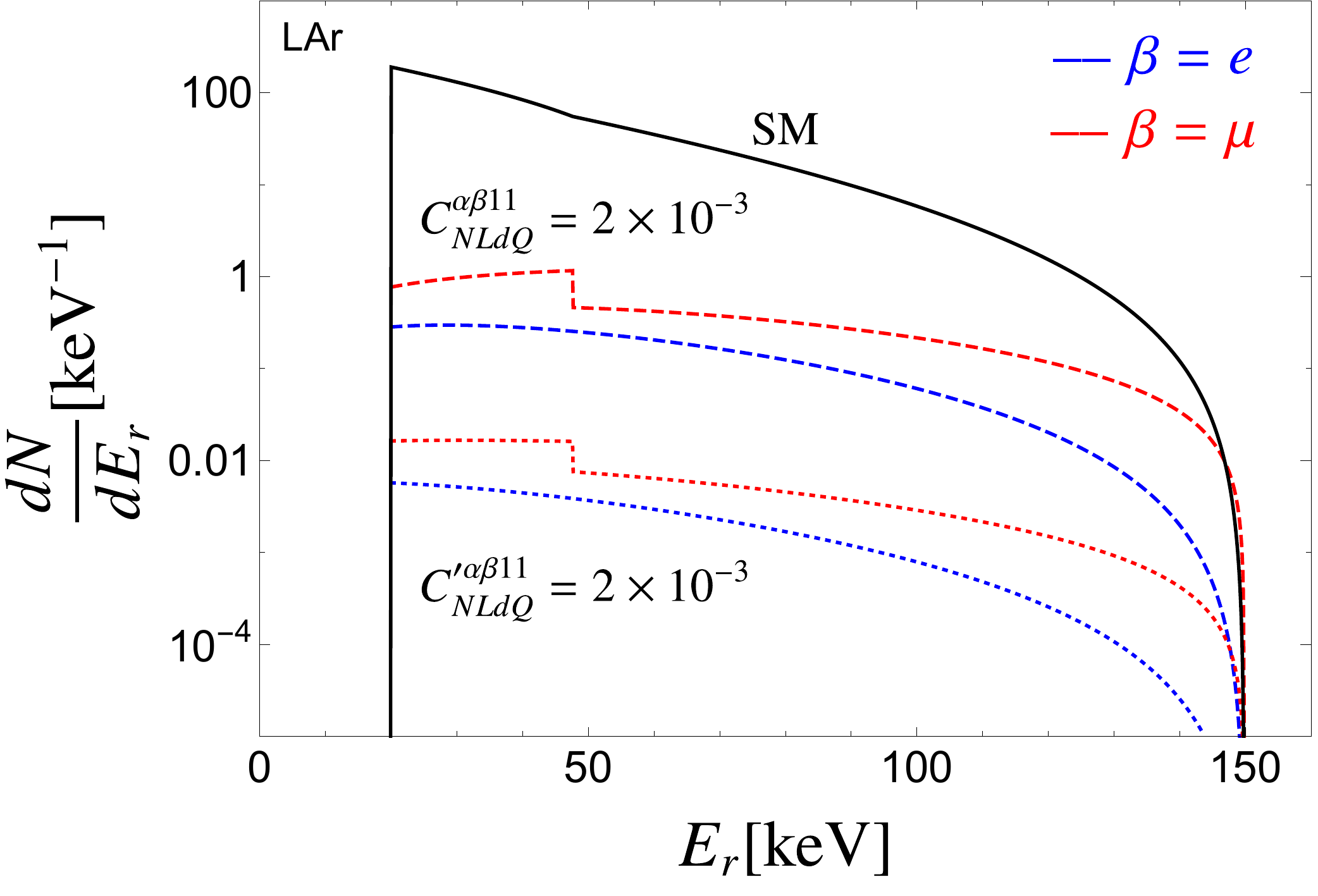}
		\caption{}
		\label{fig:CEvNEn_EFT}
	\end{subfigure}
	\begin{subfigure}{.49\textwidth}
		\includegraphics[width=\textwidth]{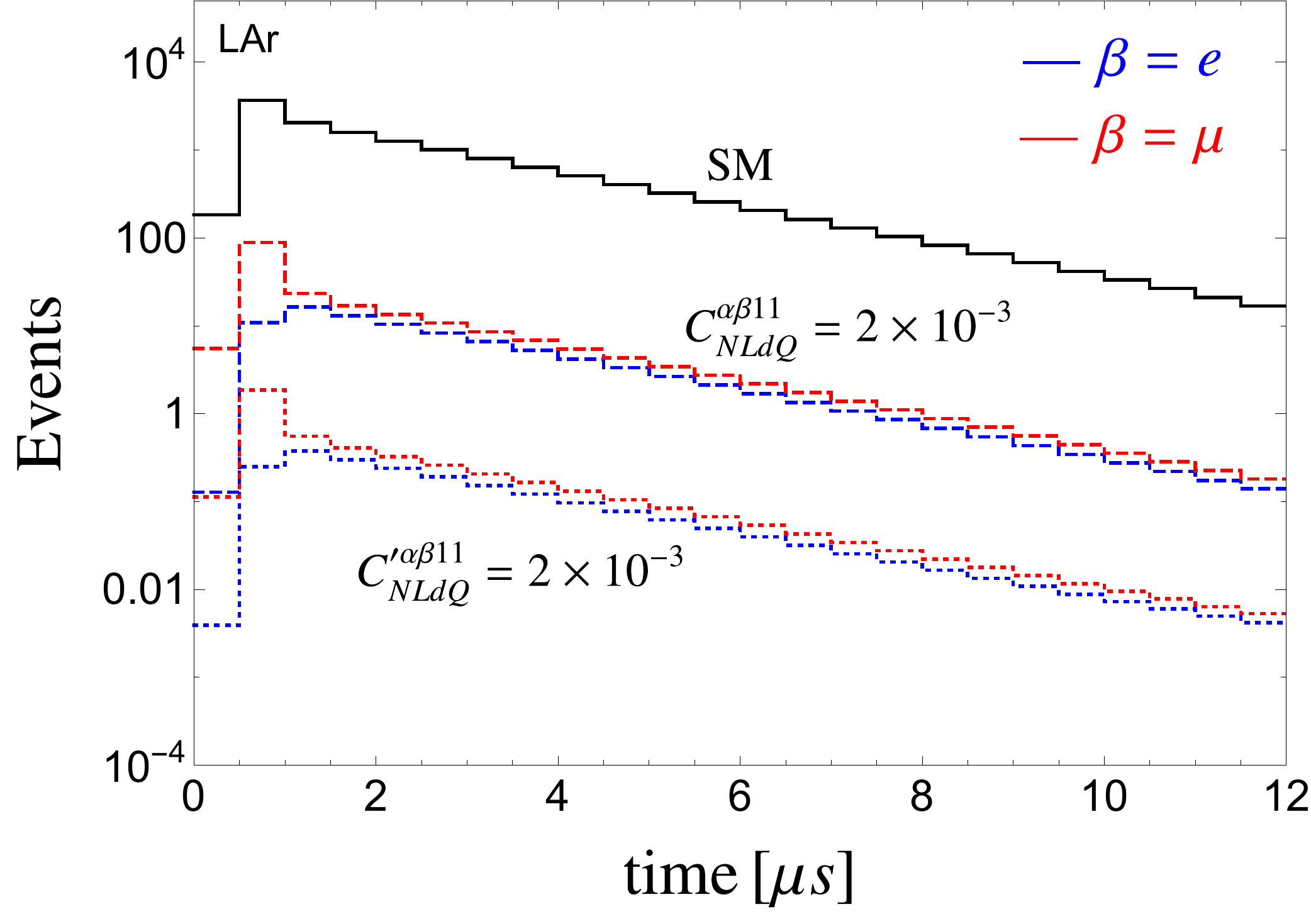}
		\caption{ }
		\label{fig:CEvNSTi_EFT}
	\end{subfigure}
	\caption{The recoil energy (left) and temporal (right) distributions in a future COHERENT LAr detector. Threshold effects are included. The black solid lines are the SM case including all flavors. The blue (red) curves correspond to the electron (muon+antimuon) flavor contributions. The dashed (dotted) curves correspond to the contributions from  the scalar (tensor) interactions with $C_{NLdQ}\, (C^\prime_{NLdQ}) = 2\times 10^{-3}$.}
	\label{fig:CEvNS_EFT}
\end{figure}

\begin{figure}
	\centering
	\includegraphics[width=0.5\columnwidth]{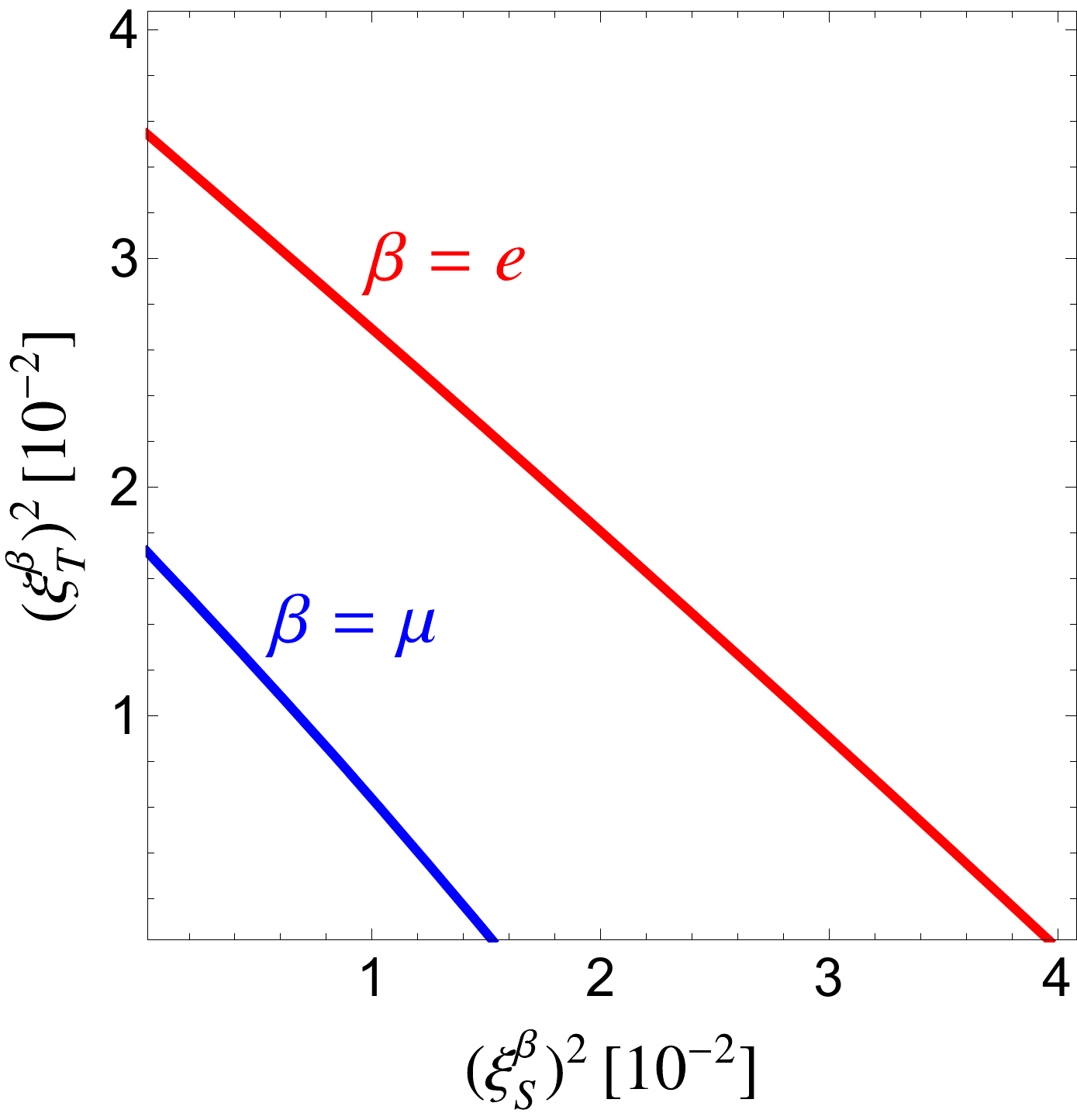}
	\caption{Projected 90\% C.L. upper bounds from the future COHERENT experiment with a 610~kg fiducial mass of LAr. }
	\label{fig:Xi}
\end{figure}

Following Ref.~\cite{Han:2019zkz}, we study the current
and projected constraints on the three GNI from the COHERENT experiment. Several COHERENT experiments with multiple targets have been proposed. In this study, we consider a future 750~kg liquid argon (LAr) detector with a 610~kg fiducial mass taking data for four years.
The energy threshold is around 20~keV, which is higher than the 6.5~keV CsI energy threshold. The observed event distributions based on the SM simulations are shown in Fig.~\ref{fig:CEvNS_dis}. The future  LAr experiment will provide much more statistics even though it has a higher threshold of nuclear recoil energy. GNI can modify the shape of the recoil energy and temporal distributions. The scalar and tensor GNI distributions compared to the SM are shown in Fig.~\ref{fig:CEvNS_EFT}. The muon flavor contributions dominate over the electron flavor as there are twice as many muon flavor neutrinos as electron flavor neutrinos. Since the $\overline{\nu}_\mu$ energy distribution peaks at the end point $m_\mu/2$, there are more events in the tail of the energy spectrum for GNI involving the muon flavor. Another observation from Fig.~\ref{fig:CEvNS_EFT} is that the COHERENT experiment is much more sensitive to the scalar operators than the tensor operator. (Note from Eq.~\ref{Eq:eps_C}  that $|\epsilon_{S,d}|$ can be larger than $|\epsilon_{T,d}|$ even if $|C_{NLdQ}|$ is smaller than $|C^\prime_{NLdQ}|$, and that the factors $m_n/m_q$ and $m_p/m_q$ in $\xi_S$ make CE$\nu$NS more sensitive to scalar GNI than tensor GNI.) 
By using the energy spectrum of the current COHERENT data, we find that the current 90\%~C.L. bounds on the scalar or tensor interactions, allowing only a single nonzero parameter, are 
\beq
(\xi^{\mu}_S)^2 < 0.60\,,\quad (\xi^{\mu}_T)^2 < 0.73\,, \quad (\xi^{e}_S)^2 < 1.5\,,\quad (\xi^{e}_T)^2 < 1.6\,.
\eeq
Also, the projected 90\%~C.L. bounds from future COHERENT data by using both the spectral and temporal information are
\beq
(\xi^{\mu}_S)^2 < 0.012\,,\quad (\xi^{\mu}_T)^2 < 0.013\,, \quad (\xi^{e}_S)^2 < 0.030\,, \quad (\xi^{e}_T)^2 < 0.027\,,
\eeq
which is  an order of magnitude improvement. Again, the bounds are set based on only one of them being nonzero.
The projected 90\%~C.L. bounds in the $(\xi^\beta_S)^2$-$(\xi^\beta_T)^2$ plane are shown in Fig.~\ref{fig:Xi}. Because of  degeneracies between the SMNEFT WCs in Eq.~(\ref{Eq:eps_C}), bounds on individual parameters cannot be placed if all the parameters are allowed to float.
The bounds on the individual  parameters can be derived after running and matching. The current (projected) 90\%~C.L. bounds on SMNEFT WCs, after setting the others to zero, are
\beq
\begin{split}
&\mid C^{\alpha e11}_{NLQu} \mid < 8.1 \times 10^{-2}\,(3.2 \times 10^{-3})\,,\quad\mid C^{\alpha\mu11}_{NLQu} \mid < 5.1 \times 10^{-2}\,(2.0 \times 10^{-3}),\\
&\mid C^{\alpha e11}_{NLdQ} \mid < 7.7 \times 10^{-2} (3.1 \times 10^{-3})\,,\quad\mid C^{\alpha\mu11}_{NLdQ} \mid <4.9 \times 10^{-2}\,(1.9 \times 10^{-3}),\\
&\mid C^{\prime \alpha e11}_{NLdQ} \mid < 2.0 \times 10^{-1}\,(2.1 \times 10^{-2})\,,\quad \mid C^{\prime\alpha\mu11}_{NLdQ} \mid < 1.4 \times 10^{-1}\,(1.4 \times 10^{-2})\,.
\end{split}
\eeq
 The projected 90\%~C.L. bounds in the $C_{NLQu}$-$C_{NLdQ}$ ($C_{NLdQ}$-$C^\prime_{NLdQ}$)  planes, are shown by the brown dashed contours in the upper (lower) panels of Fig.~\ref{fig:Coll_Scalar_EFT}. We have set $C_{NLQu}$ = 0 in the $C_{NLdQ}$-$C^\prime_{NLdQ}$ planes, because otherwise the bounds are too weak to display. The current COHERENT bounds are not shown as they are irrelevant in comparison.

\section{Collider constraints}
\label{sec:high}
\label{sec:collider}
High-energy colliders can set strong bounds on the Wilson coefficients of scalar, pseudoscalar, and tensor interactions. In this section, we study the sensitivity to the WCs at proton-proton and electron-proton colliders. We set bounds using the LHC and evaluate the potential of the HL-LHC and LHeC to probe GNI.  By integrating over the full phase space, we find the partonic cross sections of the SM mediated by the $W$ boson and of the contact scalar and tensor interactions to be~\cite{Alcaide:2019pnf}
\beqa
	\text{LHC}:  &&\hat{\sigma}_S =\frac{G_F^2 \hat{s}}{24\pi} C_S^2\,, \quad \hat{\sigma}_T =\frac{2G_F^2 \hat{s}}{9\pi} C_T^2\,,  \quad \hat{\sigma}_{SM}(u\bar d \to W^*\to \mu^+ \nu_\mu) =\frac{G_F^2 \hat{s}}{18 \pi}
	\frac{M_W^4}{(\hat{s} - M_W^2)^2} \,, ~~~~~\label{Eq:XS1}\\
	\text{LHeC}: &&\hat{\sigma}_S =\frac{G_F^2 \hat{s}}{24\pi} C_S^2\,, \quad \hat{\sigma}_T =\frac{14 G_F^2 \hat{s}}{3\pi} C_T^2\,,  \quad \hat{\sigma}_{SM}(eq\to \nu_e q') =
	\frac{G_F^2 \hat{s}}{2 \pi}
	\frac{M_W^2}{\hat{s} + M_W^2} \,, ~~
	\label{Eq:XS2}
\eeqa
where $C_S \in \{C_{NLQu}, C_{NLdQ}\}$, $C_T = C^\prime_{NLdQ}$. 
Note that the ratios of the tensor to scalar cross sections, $ \hat{\sigma}_T/\hat{\sigma}_S$, 
are $16/3$ at the LHC and $112$ at the LHeC. Clearly, the LHeC is much more sensitive to tensor interactions than scalar interactions. Owing to its lower center-of-mass energy, we expect bounds derived from the LHeC to be weaker than those from the LHC, given the $\hat s$-dependence of the higher-dimensional operators.

The interference between chirality-flipped operators and SM operators are helicity suppressed, and the interference between the scalar ($O_{NLdQ}$) and tensor ($O^\prime_{NLdQ}$) interactions is generally nonzero. 
The differential distributions for the interference of the latter operators in the center-of-mass frame are found to be
\beq
\begin{split}
&\text{LHC:}\quad \frac{d\hat{\sigma}_{ST}}{d\cos\theta^*} = \frac{G_F^2 \hat{s}}{12\pi}(C_S^* C_T+C_T^* C_S)\cos\theta^*\,, \\
&\text{LHeC:}\quad \frac{d\hat{\sigma}_{ST}}{d\cos\theta^*} = \frac{G_F^2 \hat{s}}{16\pi}(C_S^* C_T+C_T^* C_S)(\cos^2\theta^*-2\cos\theta^*-3)\,.
\end{split}
\eeq
The interference leads to a linear asymmetry at the LHC and the integrated rate vanishes, while the integrated rate at the LHeC is 
$\hat{\sigma}_{ST} = -G_F^2 \hat{s}/ {3\pi} (C_S^* C_T+C_T^* C_S).$

The hadronic cross sections can be obtained by convolving with the parton distribution functions,
\beqa
&&\sigma_{LHC} = \sum_{q,q^\prime} \int_{\tau_{min}}^{\tau_{max}} d\tau \int_{\tau}^{1} \frac{dx}{x}f_q(x,\mu_F)f_{q^\prime}(\tau/x,\mu_F)\hat{\sigma}(\tau s)\,, \\ 
%
&&\sigma_{LHeC} = \sum_{q} \int_{x_{min}}^{x_{max}} dx f_q(x,\mu_F)\hat{\sigma}(x s)\,.
\eeqa 
%
In the following, we use the Monte Carlo event generator $\rm{MadGraph5\_aMC@NLO}$~\cite{Alwall:2014hca} to generate signal and background samples at the LHC and LHeC. The GNI Lagrangian is implemented in the FeynRules 2.0~\cite{Alloul:2013bka} framework. PYTHIA8~\cite{Sjostrand:2014zea} (PYTHIA6~\cite{Sjostrand:2007gs}) is used for parton showering and hadronization at the LHC (LHeC). We perform the detector simulations using $\rm{Delphes}$ $\rm{3.4.1}$~\cite{deFavereau:2013fsa}. 

Before evaluating the collider sensitivity to the Wilson coefficients, we note that our EFT description is valid only for $\sqrt{\hat s} < \Lambda$, which calls for an assumption about the energy scale of the new physics. We consider two representative  scenarios of the new physics scale, which we call low-scale new physics (LNP) with $\Lambda \sim 1$~TeV, and high-scale new physics (HNP) with $\Lambda \gg 1$~TeV. In the HNP case, we assume the EFT method to be valid for the entire energy scale relevant to LHC data. In the LNP case, however, we limit our analysis to a subset of the LHC data below 1~TeV.

\subsection{Proton-proton colliders}
Both scalar and tensor CC contact interactions can be probed at  high-energy proton-proton colliders, under the assumption that the energy scale of the new dynamics is not kinematically accessible. The signal channel is the Drell-Yan (DY) process, $pp\rightarrow \ell \nu + X$. Due to the missing neutrino in the final state, our analysis is based on the distribution of the transverse mass, which is reconstructed by the charged lepton transverse momentum ($p_T^\ell$) and the missing transverse momentum ($E_T^{\text{miss}}$)
\beq
m_T = \sqrt{2p_T^\ell E_T^{\text{miss}}(1-\cos\Delta\phi(p_T^\ell,E_T^{\text{miss}})) }\,.
\eeq
The main background for large values of $m_T$ is DY production of $W$ bosons. The latest analysis for charged lepton and missing transverse momentum events conducted by ATLAS used 139~$\rm{fb}^{-1}$ of data collected at  $\sqrt{s} = 13$~TeV~\cite{Aad:2019wvl}. 
In the rest of our study, we only use the $m_T$ distributions below 800~GeV for the LNP scenario, and the full range of $m_T$ for the HNP scenario.
For our analyses, we define the statistical significance in terms of 
\beq
\chi^2 = \sum_{i}\frac{(n_{b,i}+n_{s,i} - n_{\text{data},i})^2}{n_{\text{data},i}+(\sigma_{i} n_{\text{data},i})^2}\,,
\label{chi2_coll}
\eeq
where $n_{b (\text{data}),i}$ is the number of background (observed) events in the $i^{th}$ bin, which is obtained directly from Ref.~\cite{Aad:2019wvl}. $n_{s,i}$ is the number of signal events simulated in Madgraph at LO. $\sigma_{i}$ is the total systematic uncertainty, which is chosen according to Ref.~\cite{Aad:2019wvl}:
\begin{equation*}
\begin{split}
&\text{electron channel: }\sigma_e \sim 10\%\, (12\%) \text{ for } m_T = 300\,(2000)\text{ GeV};\\
&\text{muon channel: }\sigma_\mu \sim 10\%\, (17\%) \text{ for } m_T = 300\,(2000)\text{ GeV}.
\end{split}
\end{equation*}
The current 90\%~C.L. bounds, defined by $\Delta\chi^2 < 2.71$,  on the LNP (HNP) scalar and tensor operators are
\beq
	\mid C^{\alpha e11}_{NLQu} \mid < 2.5\,(0.44)  \times 10^{-3}\,,  \mid C^{\alpha e11}_{NLdQ} \mid < 2.6\,(0.46) \times 10^{-3}\,, \mid C^{\prime \alpha e11}_{NLdQ} \mid < 1.2\,(0.24) \times 10^{-3}\,,
\eeq
\beq
	\mid C^{\alpha\mu11}_{NLQu} \mid < 2.9\,(0.66)  \times 10^{-3}\,, \mid C^{\alpha\mu11}_{NLdQ} \mid < 3.0\,(0.68) \times 10^{-3}\,, \mid C^{\prime\alpha\mu11}_{NLdQ} \mid < 1.4\,(0.40) \times 10^{-3}\,.
\eeq
 The bounds on $C_{NLQu}$ are slightly stronger than for $C_{NLdQ}$ because of the size of the CKM matrix element $V_{ud}$. These bounds are consistent with those in Ref.~\cite{Aad:2019wvl}. The 90\%~C.L. allowed regions in the $C_{NLdQ}$-$C_{NLQu}$ and $C_{NLdQ}$-$C^\prime_{NLdQ}$ planes are shown in the Fig.~\ref{fig:Coll_Scalar_EFT}. The solid red (blue) contours correspond to the LNP (HNP) case. 
We have checked numerically using Madgraph that the interference between scalar operator $O_{NLdQ}$ and tensor operator $O^\prime_{NLdQ}$ can be ignored.

\begin{figure}[tb]
	\centering
	\begin{subfigure}{.49\textwidth}
		\includegraphics[width=\textwidth]{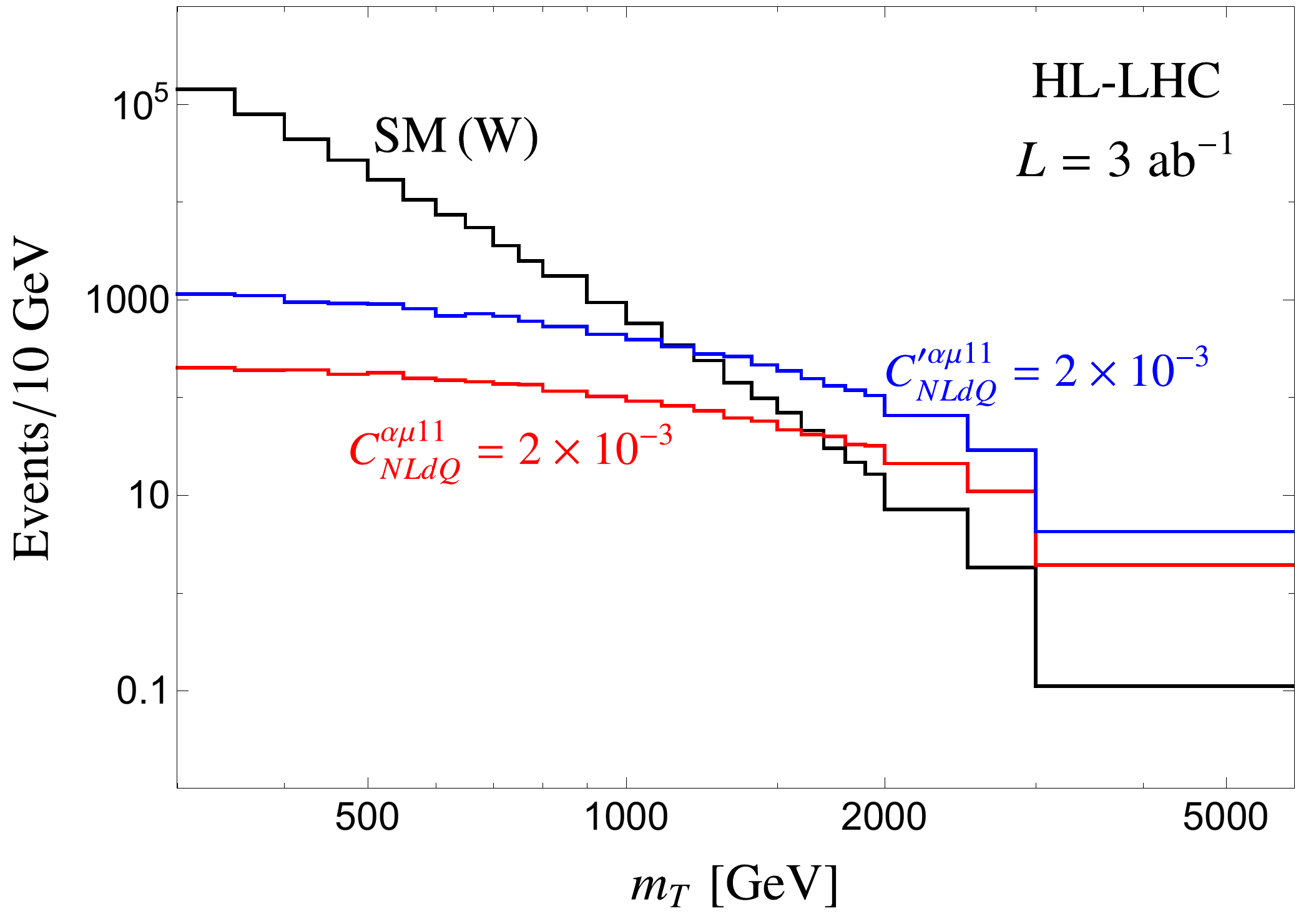}
		\caption{}
		\label{fig:mtdis}
	\end{subfigure}
	\begin{subfigure}{.49\textwidth}
		\includegraphics[width=\textwidth]{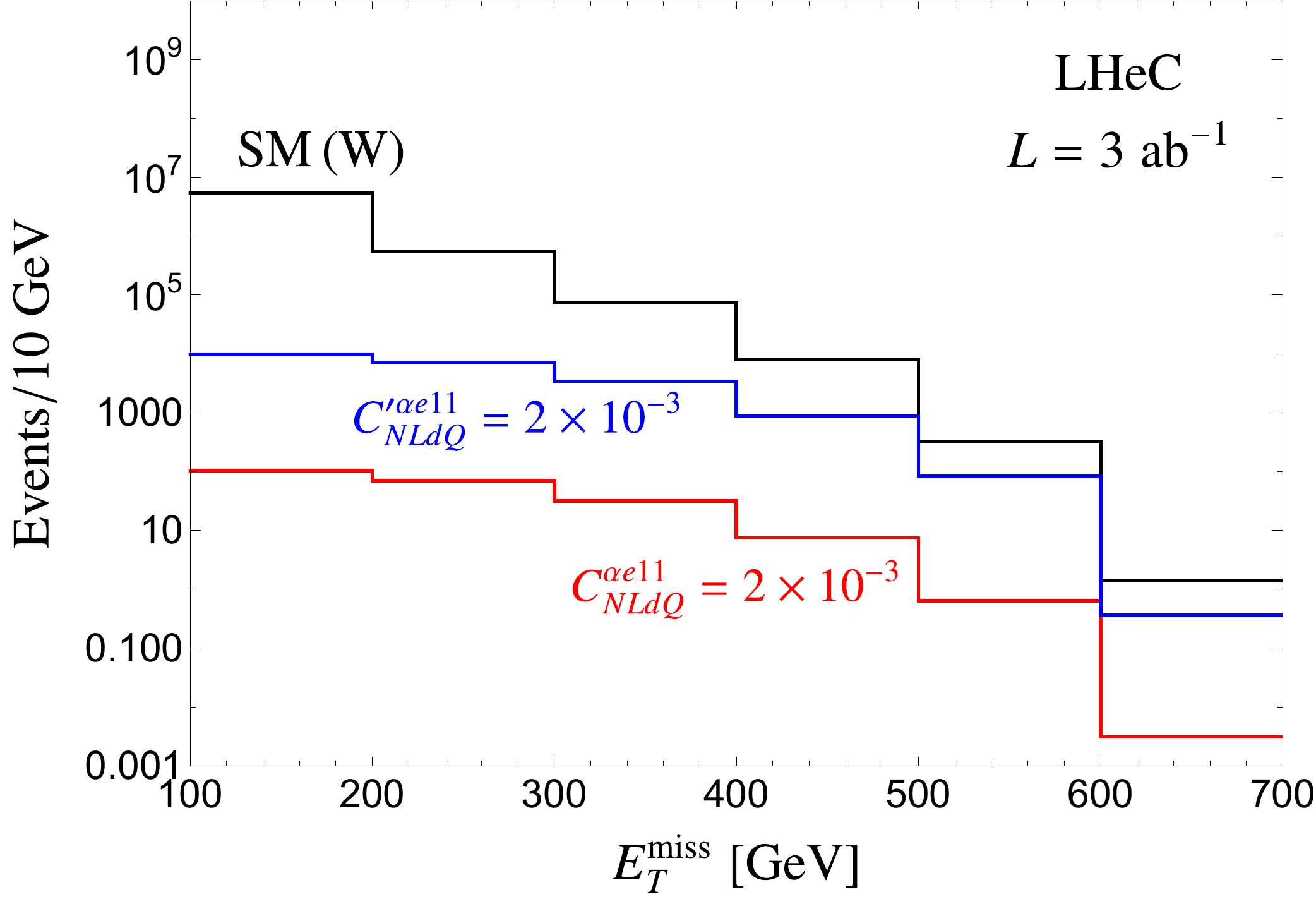}
		\caption{ }
		\label{fig:metdis}
	\end{subfigure}
	\caption{Left: Distribution of $m_T$ at the HL-LHC with an integrated luminosity of 3~$\text{ab}^{-1}$. Right: Distribution of $E^{\text{miss}}_T$  at the LHeC with 3~$\text{ab}^{-1}$ and a 1.3~TeV center-of-mass energy.   The black histograms corresponds to the SM. The red (blue) histograms correspond to  scalar (tensor) interactions with $C_{NLdQ}\, (C^\prime_{NLdQ}) = 2\times 10^{-3}$. }
\end{figure}

To assess the future potential of the LHC, we assume an integrated luminosity of $L = 3 \text{ ab}^{-1}$  and $\sqrt{s} = 14$~TeV at the HL-LHC. In this analysis, we simulate the DY $W$ background at LO multiplied by a scale factor obtained from Ref.~\cite{Aad:2019wvl}, to include other smaller backgrounds including top pairs, single top, $W\rightarrow \tau \nu$, DY $Z$, and di-bosons. The signals are also generated at tree level. We do not include a $K$ factor as it applies to both signal and background, so the significance is simply scaled by $\sqrt{K}$ after including higher-order corrections. The selection rules applied in this analysis are slightly different between the electron and muon final states. For the muon (electron) final states, we require 
\beq
\begin{split}
	&\bullet\;  p^{\mu (e)}_T > 55\,(65)\, \text{GeV and}\ |\eta_\ell| < 2.4 ,\\
	&\bullet\;  \text{veto $b$-tagged jets},\\
	&\bullet\;   \text{discard additional electron or muon with } p_T > 20\, \text{GeV and} \,|\eta_\ell| < 2.4, \\
	&\bullet\;   m_T > 300\text{ GeV},
\end{split}
\eeq
in which, the electron $p_T$ cut is slightly stronger than the muon $p_T$ cut, in order to suppress the non-prompt backgrounds. The distributions of $m_T$ above 300~GeV after applying the cuts are shown in Fig~\ref{fig:mtdis}. Deviations from the SM arise in the tails of the $m_T$ distributions because the sub-process cross sections for a dim-6 operator scale as $\hat{s}$; see Eq.~(\ref{Eq:XS1}). For the same size WC, tensor interactions have a larger cross section than scalar interactions. The $\chi^2$ used in this analysis is defined in Eq.~(\ref{chi2_coll}), with $n_{data}$ replaced by the values from SM simulations. The projected 90\%~C.L. bounds on the LNP (HNP) scalar and tensor operators are
\beq
	\mid C^{\alpha e11}_{NLQu} \mid < 2.3\,(0.28)  \times 10^{-3}\,,\mid C^{\alpha e11}_{NLdQ} \mid < 2.4\,(0.28) \times 10^{-3}\,, \mid C^{\prime \alpha e11}_{NLdQ} \mid < 1.1\,(0.18) \times 10^{-3}\,,
\eeq
\beq
  \mid C^{\alpha\mu11}_{NLQu} \mid < 2.7\,(0.28)  \times 10^{-3}\,,\mid C^{\alpha\mu11}_{NLdQ} \mid < 2.8\,(0.29) \times 10^{-3}\,,\mid C^{\prime\alpha\mu11}_{NLdQ} \mid < 1.3\,(0.18) \times 10^{-3}\,.
\eeq
The bounds from HL-LHC on scalar (tensor)  interactions with the assumption of LNP are comparable with (much stronger than) the ones we obtained for the future COHERENT experiment. The dashed red (blue) contours in Fig.~\ref{fig:Coll_Scalar_EFT} show the
90\%~C.L. projections for the HL-LHC with the LNP (HNP) assumption. The bounds on the WCs are stronger for HNP than for LNP, because the  signals in the high-energy tails of the $m_T$ distributions are not buried  in the SM background. These bounds can be converted into limits
on the effective couplings $\kappa = \sqrt{|C|}(\Lambda/v)$ for fixed values  of the new physics scale 
$\Lambda$. The 90\%~C.L. bounds on $\kappa$ are provided in Table~\ref{Table: kappa} for LNP (with $\Lambda = 1$~TeV) and HNP
 (with $\Lambda=10$~TeV). As expected, bounds on $\kappa$ are stronger in the LNP case than the HNP case. Alternatively, if we assume that 
 $\kappa \approx 1$, then HL-LHC bounds on the WCs for HNP imply a sensitivity to $\Lambda \sim 20$~TeV. This is comparable to the expected sensitivity of $W'$ searches at the HL-LHC~\cite{ATL-PHYS-PUB-2018-044}.

\begingroup
\setlength{\tabcolsep}{6pt} 
\renewcommand{\arraystretch}{1.} 
\begin{table}
	\centering
	\begin{tabular}{c | c |  c |  c | | c | c | c}
		\toprule
		\begin{tabular}{@{}c@{}} Coupling  \end{tabular}  & $\kappa^{\alpha e11}_{NLQu}$ & $\kappa^{\alpha e11}_{NLdQ}$  & $\kappa^{\prime \alpha e11}_{NLdQ}$ & $\kappa^{\alpha\mu11}_{NLQu}$ & $\kappa^{\alpha\mu11}_{NLQu}$ & $\kappa^{\prime\alpha\mu11}_{NLQu}$\\
		\midrule
		LHC: LNP (HNP)  & $0.20\, (0.85)$ & $0.21\, (0.87)$ & $0.14\, (0.63) $ & $0.22\, (1.0) $ & $0.22\, (1.1)$ & $0.15\, (0.81)$ \\
		\midrule
		HL-LHC: LNP (HNP)   & $0.19\, (0.68)$  & $0.20\, (0.68)$  & $0.13\, (0.55)$  & $0.21\, (0.68)$ & $0.22\, (0.69)$ & $0.15\, (0.55)$ \\
		\toprule
	\end{tabular}
	\caption{Current and projected 90\% C.L. bounds on the new physics coupling $\kappa$ from LHC and HL-LHC data, respectively, for the LNP ($\Lambda = 1$~TeV) and HNP ($\Lambda = 10$~TeV) cases. }
	\label{Table: kappa}
\end{table}
\endgroup

\subsection{Electron-proton colliders}
The HERA collaboration set bounds on the contact interaction $e\nu q q^\prime$ using the charged current process, $e^\pm p \rightarrow \overset{(-)}{\nu} X$, from the $Q^2$ and $x$ distributions~\cite{Cornet:1997vy}. The lower bound on the mass scale of the contact term is around 1~TeV with the strong coupling  $\sim 4\pi$. This bound can be translated to our scenario:
\beq
\mid C^{\alpha e11}_{NLQu} \mid , \mid C^{\alpha e11}_{NLdQ} \mid, \mid C^{\prime \alpha e11}_{NLdQ} \mid \lesssim  5\,,
\eeq
which is very weak compared to bounds from high-energy colliders.

Next, we consider the future $ep$ collider, LHeC, with $\sqrt{s} = 1.3$~TeV ($E_e = 60$ GeV, $E_p = 7$~TeV) and $ L = 3\text{ ab}^{-1}$. The signal channel is mono-jet, $ep\rightarrow j\nu + X$, through the $t$-channel. The main background is mediated by SM W bosons. 
For the analysis we use the following set of basic cuts:
\beq
\begin{split}
	&\bullet\;  \text{leading jet should have } p^j_T > 20\, \text{GeV and}\ |\eta_j| < 2.5 ,\\
	&\bullet\;  \text{veto any electrons with } p^e_T > 20\, \text{GeV and}\ |\eta_e| < 2.5,\\
	&\bullet\;   \text{the angular distance between jet and missing $E_T$ should be bigger than 0.4}.\nonumber
\end{split}
\eeq
 The distributions of the missing transverse energy above 100~GeV after applying the cuts are shown in Fig~\ref{fig:metdis}.
 To maximize our  $\chi^2 = S^2/B$, in which we do not include systematic uncertainties, we select the cut on the missing transverse energy as $E^{\text{miss}}_T > 300$ GeV.  
 The projected 90\%~C.L. bounds on the individual SMNEFT WCs are
 \beq
 \mid C^{\alpha e11}_{NLQu} \mid < 3.9\times 10^{-3}\,,\quad \mid C^{\alpha e11}_{NLdQ} \mid < 4.0 \times 10^{-3}\,, \quad \mid C^{\prime \alpha e11}_{NLdQ} \mid < 0.38 \times 10^{-3}\,,
 \eeq
 with only one WC taken to be nonzero. If all parameters are allowed to be nonzero, the bounds weaken slightly due to the mixing between $O_{NLdQ}$ and $O^\prime_{NLdQ}$:
  \beq
 \mid C^{\alpha e11}_{NLQu} \mid < 3.9\times 10^{-3}\,, \quad \mid C^{\alpha e11}_{NLdQ} \mid < 6.1 \times 10^{-3}\,, \quad \mid C^{\prime \alpha e11}_{NLdQ} \mid < 0.58 \times 10^{-3}\,.
 \eeq
The projected 90\%~C.L. bounds on the $C_{NLdQ}$-$C_{NLQu}$ and $C_{NLdQ}$-$C^\prime_{NLdQ}$ planes are shown in Fig.~\ref{fig:Coll_Scalar_EFT} by the purple dashed contours. Due to the smaller center-of-mass energy, the bounds on the scalar interactions from LHeC are weaker than the ones from HL-LHC. However, for tensor interactions, the bounds from LHeC are stronger than HL-LHC for the LNP case. 

\section{Summary}
\label{sec:sum}

New physics associated with neutrinos can be studied without theoretical prejudice, with allowance for scalar, pseudoscalar, vector, axial-vector and tensor interactions of neutrinos with SM fermions. If the new physics scale is much higher than the electroweak scale, it is appropriate to work in a model-independent EFT framework below the new physics scale. GNI operators below the electroweak scale are generated by EFT operators that respect the SM gauge symmetry. 

In this work we studied scalar, pseudoscalar  and tensor neutrino interactions in the framework of SMNEFT, which extends SMEFT with right-handed neutrinos. At the dimension-six level, these interactions are produced by three less constrained and phenomenologically interesting operators, namely $O_{NLQu}$, $O_{NLdQ}$, and $O^\prime_{NLdQ}$ as in Sec.~\ref{sec:SMNEFT}. Both neutral current and charged current interactions can be induced by a single operator, which can be explored in various experiments.  To compare constraints from experiments at different energy scales, we perform the RG running above and below the weak scale, and map all the bounds into the parameter space of three WCs $C_{NLQu}$, $ C_{NLdQ}$, and $ C^\prime_{NLdQ}$ at 1~TeV. We summarize the current and projected experimental bounds on the three WCs in Tables~\ref{Table: bound_now} and \ref{Table: bound_future}.  The correlations between the three operators are shown in Fig.~\ref{fig:Coll_Scalar_EFT}. Our main conclusions are:
\begin{enumerate}
	\item Neutrino mass bounds indicate that the SMNEFT operators involving the second and third families of quarks are highly constrained, while the parameter space for neutrino interactions with the first quark generation is relatively unconstrained. This conclusion, however, is model-dependent and can be evaded.
	\item Bounds on the SMNEFT WCs from low-energy probes generally suffer from degeneracies, which are induced by RG running and matching, as is evident from Eq.~(\ref{Eq:eps_C}). The high-energy probes set bounds directly on the SMNEFT WCs, and so are not subject to degeneracies.  Low-energy probes and high-energy colliders are complementary. 
	\item Charged pion decay is extremely sensitive to the LEFT pseudoscalar operators. But, there are degeneracies when the bounds are mapped into the SMNEFT WCs. With the assumption of only one nonzero operator at a time, the bounds on the electron flavor are at the $10^{-6}$ level.
	\item The strongest current bounds on the three SMNEFT operators are from LHC charged lepton $+E_T^{\text{miss}}$ searches, and are at the $10^{-4}- 10^{-3}$ level depending on the energy range of validity of the EFT. 	
	\item HL-LHC can improve the bounds by a factor of a few  and reach $10^{-4}$ in the HNP case. For LNP, the improvement is minor because systematic uncertainties dominate for low $m_T$.
	\item Current LHC data can exclude $\kappa \gtrsim 0.14$ for $\Lambda = 1$~TeV and $\kappa \gtrsim 0.63$ for $\Lambda = 10$~TeV. Future HL-LHC data can exclude $\kappa \gtrsim 0.13$ for $\Lambda = 1$~TeV and $\kappa \gtrsim 0.55$ for $\Lambda = 10$~TeV. For strong interactions with $\kappa = 4\pi$, the new physics scale can be excluded up to 200~TeV.
	\item A future COHERENT experiment with LAr can set strong bounds on the scalar operators, comparable with that from the HL-LHC with the LNP assumption, especially when the muon flavor is involved.
	\item LHeC will be important to study tensor interactions involving the electron flavor, and can place bounds at the $10^{-4}$ level.
\end{enumerate}

\begingroup
\setlength{\tabcolsep}{10pt} 
\renewcommand{\arraystretch}{1.} 
\begin{table}
	\centering
	\begin{tabular}{c |  c|  c|  c| c | c | c}
		\toprule
		WC  &  $\pi^+$ decay  &  $\beta$ decay   &$\nu$ DIS & CE$\nu $NS &HERA  & LHC: LNP(HNP)\\
		\midrule
		$C^{\alpha e11}_{NLQu}$  & $3.3\times 10^{-6}$ & $3.4\times 10^{-2}$ & $0.77$ & $8.1\times 10^{-2}$& $\mathbf{ \sim 5}$ & $\mathbf{2.5\, (0.44)\times 10^{-3}}$\\
		\midrule
		$C^{\alpha e11}_{NLdQ}$   & $3.4\times 10^{-6}$ & $3.5\times 10^{-2}$ &$0.75$&$7.7\times 10^{-2}$&$\mathbf{ \sim 5}$ & $\mathbf{2.6\, (0.46)\times 10^{-3}}$ \\
		\midrule
		$C^{\prime \alpha e11}_{NLdQ}$   &$3.9\times 10^{-5}$& $2.8\times 10^{-2}$ & $0.15$&$0.20$& $\mathbf{\sim 5}$ & $\mathbf{1.2\, (0.24)\times 10^{-3}}$\\
		\toprule
		\toprule
		$C^{\alpha\mu11}_{NLQu}$ & $1.5\times 10^{-3}$ & - & $\mathbf{7.8\times 10^{-2}}$&$5.1\times 10^{-2}$&-  & $\mathbf{2.9\, (0.66)\times 10^{-3}}$\\
		\midrule
		$C^{\alpha\mu11}_{NLdQ}$ & $1.5\times 10^{-3}$ & - & $\mathbf{7.6\times 10^{-2}}$ &$4.9\times 10^{-2}$ &-& $\mathbf{3.0\, (0.68)\times 10^{-3}}$ \\
		\midrule
		$C^{\prime\alpha\mu11}_{NLdQ}$ & $1.7\times 10^{-2}$ & - & $\mathbf{1.5\times 10^{-2}}$&$0.14$&-& $\mathbf{1.4\, (0.40)\times 10^{-3}}$ \\
		\toprule
	\end{tabular}
	\caption{Current 90\%~C.L. bounds on the three SMNEFT WCs $C_{NLQu}$, $C_{NLdQ}$, and $C^{\prime}_{NLdQ}$, for the electron and muon flavors  at a 1~TeV energy scale. The constraints obtained by allowing all WCs to simultaneously vary are in boldface.}
	\label{Table: bound_now}
\end{table}
\endgroup

\begingroup
\setlength{\tabcolsep}{10pt} 
\renewcommand{\arraystretch}{1.} 
\begin{table}
	\centering
	\begin{tabular}{c |  c|  c|  c}
		\toprule
		WC    & CE$\nu $NS-LAr & LHeC  & HL-LHC: LNP(HNP)\\
		\midrule
		$C^{\alpha e11}_{NLQu}$  & $3.2\times 10^{-3}$ & $\mathbf{3.9\times 10^{-3}}$ & $\mathbf{2.3\,(0.28)\times 10^{-3}}$ \\
		\midrule
		$C^{\alpha e11}_{NLdQ}$   & $3.1\times 10^{-3}$ & $\mathbf{6.1\times 10^{-3}}$ &$\mathbf{2.4\,(0.28)\times 10^{-3}}$ \\
		\midrule
		$C^{\prime \alpha e11}_{NLdQ}$   &$2.1\times 10^{-2}$& $\mathbf{0.58\times 10^{-3}}$ & $\mathbf{1.1\,(0.18)\times 10^{-3}}$\\
		\toprule
		\toprule
		$C^{\alpha\mu11}_{NLQu}$ & $2.0\times 10^{-3}$ & - & $\mathbf{2.7\,(0.28)\times 10^{-3}}$\\
		\midrule
		$C^{\alpha\mu11}_{NLdQ}$ & $1.9\times 10^{-3}$ & - & $\mathbf{2.8\,(0.29)\times 10^{-3}}$ \\
		\midrule
		$C^{\prime\alpha\mu11}_{NLdQ}$ & $1.4\times 10^{-2}$ & - & $\mathbf{1.3\,(0.18)\times 10^{-3}}$ \\
		\toprule
	\end{tabular}
	\caption{Projected 90\%~C.L. bounds on the three SMNEFT WCs $C_{NLQu}$, $C_{NLdQ}$, and $C^{\prime}_{NLdQ}$, with electron and muon flavor,  at 1 TeV energy scale. The constraints obtained by allowing all WCs to simultaneously vary are in boldface. }
	\label{Table: bound_future}
\end{table}
\endgroup

\acknowledgments
We thank Keping Xie for helpful discussions. The work of T.H.~and H.L.~was supported in part by the U.S.~Department of Energy under grant No.~DE-FG02-95ER40896 and in part by the PITT PACC. The work of J.L. was supported by the National Natural Science Foundation of China under Grant No.~11905299. D.M. was supported in part by the U.S. Department of Energy
under Grant No.~de-sc0010504.

\newpage
\begin{appendix}
	\section{1-loop RG running}
	\label{RGrunning}

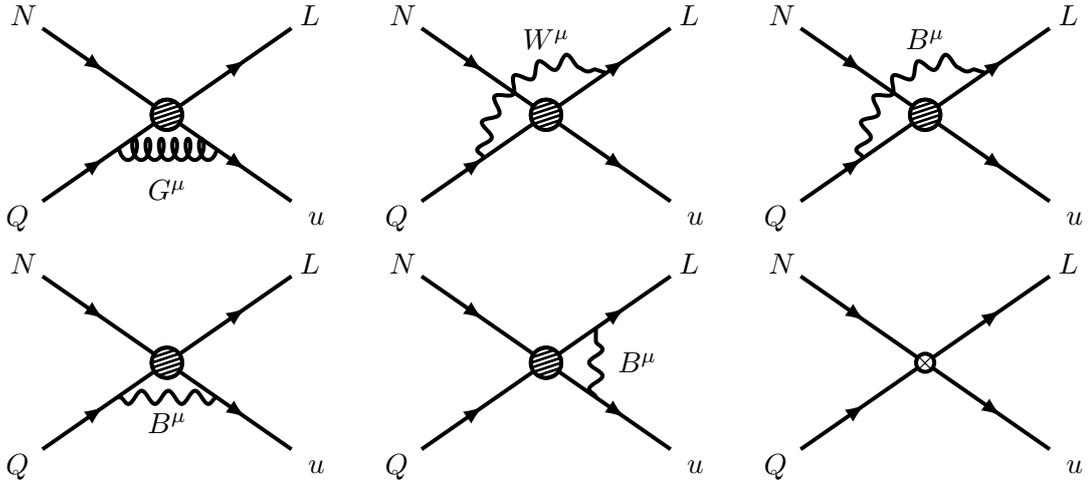
\begin{figure}
	\centering
	\begin{subfigure}{.3\textwidth}
		\begin{tikzpicture}[line width=1.5 pt, scale=2]
		\draw[fermion](145:1) -- (145:.1cm);
		\node at (145:1.15) {$N$};	
		\draw[fermion](215:1) -- (215:.1cm);
		\node at (215:1.2) {$Q$};	
		\draw[fermionbar](35:1) -- (35:.1cm);
		\node at (35:1.15) {$L$};	
		\draw[gluon] (-35:0.4) -- (215:.4cm);
		\node at (90:-.5) {$G^\mu$};	
		\draw[fermionbar](-35:1) -- (-35:.1cm);
		\node at (-35:1.2) {$u$};	
		\draw[fill=black] (0,0) circle (.1cm);
		\draw[fill=white] (0,0) circle (.1cm);
		\begin{scope}
		\clip (0,0) circle (.1cm);
		\foreach \x in {-.9,-.8,...,.3}
		\draw[line width=1 pt] (\x,-.1) -- (\x+.6,.1);
		\end{scope}
		\end{tikzpicture}	
	\end{subfigure}
	\begin{subfigure}{.3\textwidth}
		\begin{tikzpicture}[line width=1.5 pt, scale=2]
		\draw[fermion](145:1) -- (145:.1cm);
		\node at (145:1.15) {$N$};
		\draw[fermion](215:1) -- (215:.1cm);
		\node at (215:1.2) {$Q$};
		\draw[fermionbar](35:1) -- (35:.1cm);
		\node at (35:1.15) {$L$};
		\draw[vector] (215:0.5) arc (180:70:.6);
		\node at (90:.5) {$ W^\mu$};
		\draw[fermionbar](-35:1) -- (-35:.1cm);
		\node at (-35:1.2) {$u$};
		\draw[fill=black] (0,0) circle (.1cm);
		\draw[fill=white] (0,0) circle (.1cm);
		\begin{scope}
		\clip (0,0) circle (.1cm);
		\foreach \x in {-.9,-.8,...,.3}
		\draw[line width=1 pt] (\x,-.1) -- (\x+.6,.1);
		\end{scope}
		\end{tikzpicture}	
	\end{subfigure}
	\begin{subfigure}{.3\textwidth}
	\begin{tikzpicture}[line width=1.5 pt, scale=2]
	\draw[fermion](145:1) -- (145:.1cm);
	\node at (145:1.15) {$N$};
	\draw[fermion](215:1) -- (215:.1cm);
	\node at (215:1.2) {$Q$};
	\draw[fermionbar](35:1) -- (35:.1cm);
	\node at (35:1.15) {$L$};
	\draw[vector] (215:0.5) arc (180:70:.6);
	\node at (90:.5) {$B^\mu$};
	\draw[fermionbar](-35:1) -- (-35:.1cm);
	\node at (-35:1.2) {$u$};
	\draw[fill=black] (0,0) circle (.1cm);
	\draw[fill=white] (0,0) circle (.1cm);
	\begin{scope}
	\clip (0,0) circle (.1cm);
	\foreach \x in {-.9,-.8,...,.3}
	\draw[line width=1 pt] (\x,-.1) -- (\x+.6,.1);
	\end{scope}
	\end{tikzpicture}	
\end{subfigure}
	\begin{subfigure}{.3\textwidth}
	\begin{tikzpicture}[line width=1.5 pt, scale=2]
	\draw[fermion](145:1) -- (145:.1cm);
	\node at (145:1.15) {$N$};
	\draw[fermion](215:1) -- (215:.1cm);
	\node at (215:1.2) {$Q$};
	\draw[fermionbar](35:1) -- (35:.1cm);
	\node at (35:1.15) {$L$};
	\draw[vector] (-35:0.4) -- (215:.4cm);
	\node at (90:-.4) {$B^\mu$};
	\draw[fermionbar](-35:1) -- (-35:.1cm);
	\node at (-35:1.2) {$u$};
	\draw[fill=black] (0,0) circle (.1cm);
	\draw[fill=white] (0,0) circle (.1cm);
	\begin{scope}
	\clip (0,0) circle (.1cm);
	\foreach \x in {-.9,-.8,...,.3}	
	\draw[line width=1 pt] (\x,-.1) -- (\x+.6,.1);
	\end{scope}
	\end{tikzpicture}	
\end{subfigure}
	\begin{subfigure}{.3\textwidth}
		\begin{tikzpicture}[line width=1.5 pt, scale=2]
		\draw[fermion](145:1) -- (145:.1cm);
		\node at (145:1.15) {$N$};
		\draw[fermion](215:1) -- (215:.1cm);
		\node at (215:1.2) {$Q$};
		\draw[fermionbar](35:1) -- (35:.1cm);
		\node at (35:1.15) {$L$};
		\draw[vector] (-35:0.4) -- (35:.4cm);
		\node at (180:-0.6) {$B^\mu$};
		\draw[fermionbar](-35:1) -- (-35:.1cm);
		\node at (-35:1.2) {$u$};
		\draw[fill=black] (0,0) circle (.1cm);
		\draw[fill=white] (0,0) circle (.1cm);
		\begin{scope}
		\clip (0,0) circle (.1cm);
		\foreach \x in {-.9,-.8,...,.3}
		\draw[line width=1 pt] (\x,-.1) -- (\x+.6,.1);
		\end{scope}
		\end{tikzpicture}
	\end{subfigure}
	\begin{subfigure}{.3\textwidth}
		\begin{tikzpicture}[line width=1.5 pt, scale=2]
		\draw[fermion](145:1) -- (145:.06cm);
		\node at (145:1.15) {$N$};
		\draw[fermion](215:1) -- (215:.06cm);
		\node at (215:1.2) {$Q$};
		\draw[fermionbar](35:1) -- (35:.06cm);
		\node at (35:1.15) {$L$};
		\draw[fermionbar](-35:1) -- (-35:.06cm);
		\node at (-35:1.2) {$u$};
		\draw[fill=white] (0,0) circle (.06cm);
		\node at (0,0) {$\times$};
		\end{tikzpicture}
	\end{subfigure}
	\caption{One-loop vertex corrections and counterterm for $O_{NLQu}$ above the weak scale. The diagrams are similar for the other two operators.}
	\label{fig:feyn_C}
\end{figure}

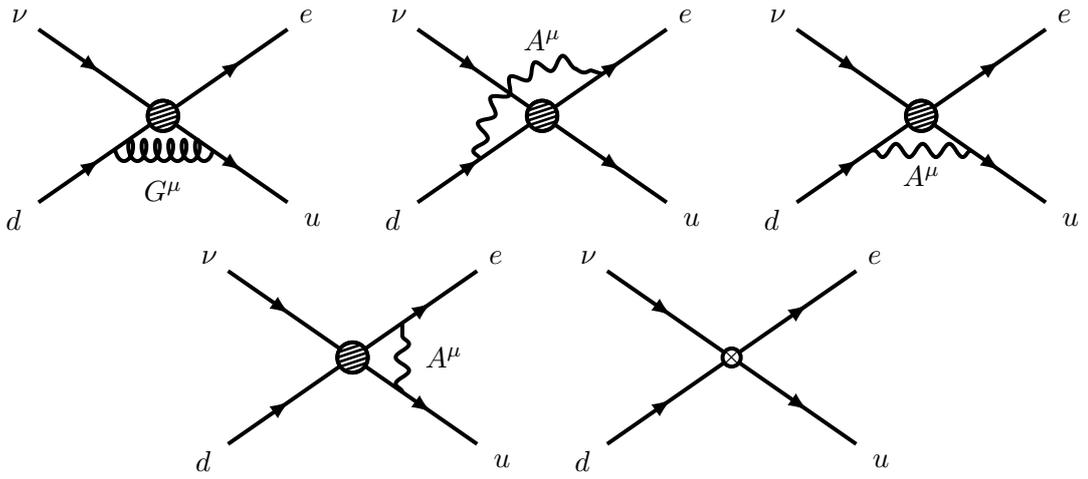
\begin{figure}
	\centering
	\begin{subfigure}{.3\textwidth}
		\begin{tikzpicture}[line width=1.5 pt, scale=2]
		\draw[fermion](145:1) -- (145:.1cm);
		\node at (145:1.15) {$\nu$};	
		\draw[fermion](215:1) -- (215:.1cm);
		\node at (215:1.2) {$d$};	
		\draw[fermionbar](35:1) -- (35:.1cm);
		\node at (35:1.15) {$e$};	
		\draw[gluon] (-35:0.4) -- (215:.4cm);
		\node at (90:-.5) {$G^\mu$};	
		\draw[fermionbar](-35:1) -- (-35:.1cm);
		\node at (-35:1.2) {$u$};	
		\draw[fill=black] (0,0) circle (.1cm);
		\draw[fill=white] (0,0) circle (.1cm);
		\begin{scope}
		\clip (0,0) circle (.1cm);
		\foreach \x in {-.9,-.8,...,.3}
		\draw[line width=1 pt] (\x,-.1) -- (\x+.6,.1);
		\end{scope}
		\end{tikzpicture}	
	\end{subfigure}
	\begin{subfigure}{.3\textwidth}
		\begin{tikzpicture}[line width=1.5 pt, scale=2]
		\draw[fermion](145:1) -- (145:.1cm);
		\node at (145:1.15) {$\nu$};
		\draw[fermion](215:1) -- (215:.1cm);
		\node at (215:1.2) {$d$};
		\draw[fermionbar](35:1) -- (35:.1cm);
		\node at (35:1.15) {$e$};
		\draw[vector] (215:0.5) arc (180:70:.6);
		\node at (90:.5) {$ A^\mu$};
		\draw[fermionbar](-35:1) -- (-35:.1cm);
		\node at (-35:1.2) {$u$};
		\draw[fill=black] (0,0) circle (.1cm);
		\draw[fill=white] (0,0) circle (.1cm);
		\begin{scope}
		\clip (0,0) circle (.1cm);
		\foreach \x in {-.9,-.8,...,.3}
		\draw[line width=1 pt] (\x,-.1) -- (\x+.6,.1);
		\end{scope}
		\end{tikzpicture}	
	\end{subfigure}
	\begin{subfigure}{.3\textwidth}
		\begin{tikzpicture}[line width=1.5 pt, scale=2]
		\draw[fermion](145:1) -- (145:.1cm);
		\node at (145:1.15) {$\nu$};
		\draw[fermion](215:1) -- (215:.1cm);
		\node at (215:1.2) {$d$};
		\draw[fermionbar](35:1) -- (35:.1cm);
		\node at (35:1.15) {$e$};
		\draw[vector] (-35:0.4) -- (215:.4cm);
		\node at (90:-.4) {$A^\mu$};
		\draw[fermionbar](-35:1) -- (-35:.1cm);
		\node at (-35:1.2) {$u$};
		\draw[fill=black] (0,0) circle (.1cm);
		\draw[fill=white] (0,0) circle (.1cm);
		\begin{scope}
		\clip (0,0) circle (.1cm);
		\foreach \x in {-.9,-.8,...,.3}	
		\draw[line width=1 pt] (\x,-.1) -- (\x+.6,.1);
		\end{scope}
		\end{tikzpicture}	
	\end{subfigure}
	\begin{subfigure}{.3\textwidth}
		\begin{tikzpicture}[line width=1.5 pt, scale=2]
		\draw[fermion](145:1) -- (145:.1cm);
		\node at (145:1.15) {$\nu$};
		\draw[fermion](215:1) -- (215:.1cm);
		\node at (215:1.2) {$d$};
		\draw[fermionbar](35:1) -- (35:.1cm);
		\node at (35:1.15) {$e$};
		\draw[vector] (-35:0.4) -- (35:.4cm);
		\node at (180:-0.6) {$A^\mu$};
		\draw[fermionbar](-35:1) -- (-35:.1cm);
		\node at (-35:1.2) {$u$};
		\draw[fill=black] (0,0) circle (.1cm);
		\draw[fill=white] (0,0) circle (.1cm);
		\begin{scope}
		\clip (0,0) circle (.1cm);
		\foreach \x in {-.9,-.8,...,.3}
		\draw[line width=1 pt] (\x,-.1) -- (\x+.6,.1);
		\end{scope}
		\end{tikzpicture}
	\end{subfigure}
	\begin{subfigure}{.3\textwidth}
		\begin{tikzpicture}[line width=1.5 pt, scale=2]
		\draw[fermion](145:1) -- (145:.06cm);
		\node at (145:1.15) {$\nu$};
		\draw[fermion](215:1) -- (215:.06cm);
		\node at (215:1.2) {$d$};
		\draw[fermionbar](35:1) -- (35:.06cm);
		\node at (35:1.15) {$e$};
		\draw[fermionbar](-35:1) -- (-35:.06cm);
		\node at (-35:1.2) {$u$};
		\draw[fill=white] (0,0) circle (.06cm);
		\node at (0,0) {$\times$};
		\end{tikzpicture}
	\end{subfigure}
	\caption{One-loop vertex corrections and counterterm for CC operators.}
	\label{fig:feyn_eud}
\end{figure}

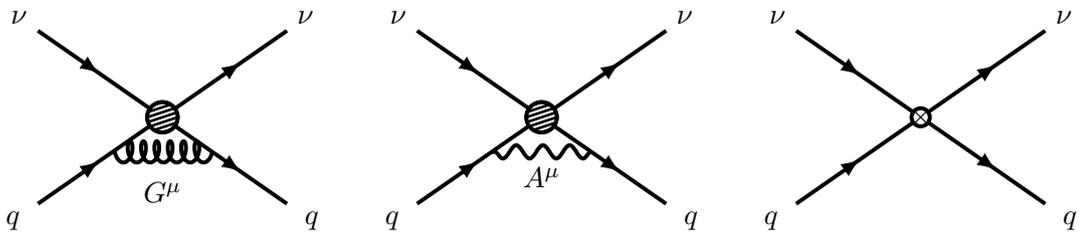
\begin{figure}
	\centering
	\begin{subfigure}{.3\textwidth}
		\begin{tikzpicture}[line width=1.5 pt, scale=2]
		\draw[fermion](145:1) -- (145:.1cm);
		\node at (145:1.15) {$\nu$};	
		\draw[fermion](215:1) -- (215:.1cm);
		\node at (215:1.2) {$q$};	
		\draw[fermionbar](35:1) -- (35:.1cm);
		\node at (35:1.15) {$\nu$};	
		\draw[gluon] (-35:0.4) -- (215:.4cm);
		\node at (90:-.5) {$G^\mu$};	
		\draw[fermionbar](-35:1) -- (-35:.1cm);
		\node at (-35:1.2) {$q$};	
		\draw[fill=black] (0,0) circle (.1cm);
		\draw[fill=white] (0,0) circle (.1cm);
		\begin{scope}
		\clip (0,0) circle (.1cm);
		\foreach \x in {-.9,-.8,...,.3}
		\draw[line width=1 pt] (\x,-.1) -- (\x+.6,.1);
		\end{scope}
		\end{tikzpicture}	
	\end{subfigure}
	\begin{subfigure}{.3\textwidth}
		\begin{tikzpicture}[line width=1.5 pt, scale=2]
		\draw[fermion](145:1) -- (145:.1cm);
		\node at (145:1.15) {$\nu$};
		\draw[fermion](215:1) -- (215:.1cm);
		\node at (215:1.2) {$q$};
		\draw[fermionbar](35:1) -- (35:.1cm);
		\node at (35:1.15) {$\nu$};
		\draw[vector] (-35:0.4) -- (215:.4cm);
		\node at (90:-.4) {$A^\mu$};
		\draw[fermionbar](-35:1) -- (-35:.1cm);
		\node at (-35:1.2) {$q$};
		\draw[fill=black] (0,0) circle (.1cm);
		\draw[fill=white] (0,0) circle (.1cm);
		\begin{scope}
		\clip (0,0) circle (.1cm);
		\foreach \x in {-.9,-.8,...,.3}	
		\draw[line width=1 pt] (\x,-.1) -- (\x+.6,.1);
		\end{scope}
		\end{tikzpicture}	
	\end{subfigure}
	\begin{subfigure}{.3\textwidth}
		\begin{tikzpicture}[line width=1.5 pt, scale=2]
		\draw[fermion](145:1) -- (145:.06cm);
		\node at (145:1.15) {$\nu$};
		\draw[fermion](215:1) -- (215:.06cm);
		\node at (215:1.2) {$q$};
		\draw[fermionbar](35:1) -- (35:.06cm);
		\node at (35:1.15) {$\nu$};
		\draw[fermionbar](-35:1) -- (-35:.06cm);
		\node at (-35:1.2) {$q$};
		\draw[fill=white] (0,0) circle (.06cm);
		\node at (0,0) {$\times$};
		\end{tikzpicture}
	\end{subfigure}
	\caption{One-loop vertex corrections and counterterm for NC operators.}
	\label{fig:feyn_eq}
\end{figure}

We define the SMNEFT Lagrangian as
\beq
\lag_{\text{SMNEFT}}\supset \sum_{i=1}^3 C_i O_i\,,
\eeq
where $\vec{O}\equiv \{O_{NLQu},O_{NLdQ},O^\prime_{NLdQ}\}$ and $\vec{C}\equiv 2\sqrt{2} G_F \{C_{NLQu},C_{NLdQ},C^\prime_{NLdQ}\}$. The anomalous dimension of $O_i$ can be obtained by the operator renormalization,
\beq
\lag_{\text{SMNEFT}}\supset \sum_{i=1}^3 C_i \frac{Z_i}{Z_{4\psi,i}} O_i^{(0)}\,,
\eeq
where $O^{(0)}$ is the operator in terms of bare fields. $Z_{4\psi,i}$ are the four-fermion field strength renormalizations obtained from the self-energy diagrams:
\beqa
Z_{4\psi,1} &=& \sqrt{Z^L_{2L}Z^Y_{2L}}\sqrt{Z^C_{2Q}Z^L_{2Q}Z^Y_{2Q}}\sqrt{Z^C_{2u}Z^Y_{2u}}\,,\\
Z_{4\psi,2}=Z_{4\psi,3} &=& \sqrt{Z^L_{2L}Z^Y_{2L}}\sqrt{Z^C_{2Q}Z^L_{2Q}Z^Y_{2Q}}\sqrt{Z^C_{2d}Z^Y_{2d}}\,,
\eeqa
where the fermion field strengths are
\beq
Z^C_{2\psi} = 1-\frac{1}{\epsilon} \frac{2\alpha_3}{3\pi}\,,\quad Z^L_{2\psi} = 1-\frac{1}{\epsilon} \frac{3\alpha_2}{8\pi}\,, \quad Z^Y_{2\psi} = 1-Y_\psi^2\frac{1}{\epsilon} \frac{\alpha_1}{2\pi}\,.
\eeq
Here $\alpha_i \equiv \frac{g_i^2}{4\pi}$, and $Y_\psi$ is hypercharge of the fermion. $Z_i$ are the corrections from counterterms which cancels the UV divergence from the 1-loop Feynman diagrams shown in Fig.~\ref{fig:feyn_C}. We have used FeynCalc~\cite{Shtabovenko:2016sxi,Mertig:1990an} to extract the UV divergence from the one-loop integrals. The expressions for $Z_i$ are
\beqa
&&Z_1 = 1-\frac{1}{\epsilon}[\frac{8\alpha_3}{3\pi} + \frac{3\alpha_2}{8\pi}+\frac{\alpha_1}{2\pi}(Y_u (4Y_Q-Y_L)+Y_LY_Q)]\,, \\
&&Z_2 = 1-\frac{1}{\epsilon}[\frac{8\alpha_3}{3\pi} + \frac{3\alpha_2}{8\pi}(1-12\frac{C_3}{C_2})+ \frac{\alpha_1}{2\pi}(Y_L(Y_d-Y_Q)+4Y_dY_Q-12\frac{C_3}{C_2})]\,, \\
&&Z_3 = 1-\frac{1}{\epsilon}[\frac{9\alpha_2}{8\pi}(1-\frac{1}{12}\frac{C_2}{C_3})+\frac{3\alpha_1}{2\pi}(Y_L(Y_d-Y_Q)+\frac{1}{12}\frac{C_2}{C_3} Y_L(Y_d+Y_Q))]\,.
\eeqa
$Z_2$ and $Z_3$ can introduce mixing between $C_2$ and $C_3$ from the weak running. The renormalization group equations of the WCs then arise from the fact that the bare operators and Lagrangian are independent of the renormalization scale $\mu$,
\beq
\mu\frac{dC_i}{d\mu} = \gamma_{ij}C_j\,.
\eeq
It is then straightforward to calculate the anomalous dimension matrix $\gamma_{ij}$ which yield the results shown in Eq.~(\ref{eq:adm_c}). Following a similar procedure, the anomalous dimension matrix for the CC and NC WCs below the weak scale are obtained by evaluating the one-loop vertex corrections and counterterms in Figs.~\ref{fig:feyn_eud} and~\ref{fig:feyn_eq}, respectively. 

\end{appendix}

\newpage
\bibliographystyle{JHEP}
\bibliography{ref}

\end{document}